\documentclass[a4paper,11pt]{article}
\pdfoutput=1 

\usepackage{jheppub} 

\usepackage[T1]{fontenc} 
\usepackage[percent]{overpic}

\title{\boldmath The $0\nu2\beta$-decay CROSS experiment: preliminary results and prospects}

\author[a]{I.C.~Bandac,}
\author[b]{A.S.~Barabash,}
\author[c]{L.~Berg\'e,}
\author[d]{M.~Bri\`ere,}
\author[d]{Ch.~Bourgeois,}
\author[e]{P.~Carniti,}
\author[c]{M.~Chapellier,}
\author[f]{M. de Combarieu,}
\author[g]{I.~Dafinei,}
\author[h]{F.A.~Danevich,}
\author[c]{N.~Dosme,}
\author[d]{D.~Doullet,}
\author[c]{L.~Dumoulin,}
\author[i]{F.~Ferri,}
\author[c,j,1]{A.~Giuliani,\note{Corresponding author.}}
\author[e]{C.~Gotti,}
\author[i]{Ph.~Gras,}
\author[d]{E.~Guerard,}
\author[k]{A.~Ianni,}
\author[c]{H.~Khalife,}
\author[b]{S.I.~Konovalov,}
\author[c]{E.~Legay,}
\author[d]{P.~Loaiza,}
\author[c]{P.~de~Marcillac,}
\author[c]{S.~Marnieros,}
\author[c]{C.A.~Marrache-Kikuchi,}
\author[i]{C.~Nones,}
\author[c]{V.~Novati,}
\author[c]{E.~Olivieri,}
\author[c]{Ch.~Oriol,}
\author[e]{G.~Pessina,}
\author[c]{D.V.~Poda,}
\author[c]{Th.~Redon,}
\author[h]{V.I.~Tretyak,}
\author[b]{V.I.~Umatov,}
\author[h]{M.M.~Zarytsky,}
\author[c]{A.S.~Zolotarova}

\affiliation[a]{Laboratorio Subterr\'aneo de Canfranc, 22880 Canfranc-Estaci\'on, Spain}
\affiliation[b]{National Research Centre Kurchatov Institute, Institute of Theoretical and Experimental Physics, 117218 Moscow, Russia}
\affiliation[c]{CSNSM, Univ. Paris-Sud, CNRS/IN2P3, Universit\'e Paris-Saclay, F-91405 Orsay, France}
\affiliation[d]{LAL, Univ. Paris-Sud, CNRS/IN2P3, Universit\'e Paris-Saclay, F-91405 Orsay, France}
\affiliation[e]{INFN, Sezione di Milano Bicocca, I-20126 Milano, Italy}
\affiliation[f]{IRAMIS, CEA, Universit\'e Paris-Saclay, F-91191 Gif-sur-Yvette, France}
\affiliation[g]{INFN, Sezione di Roma, I-00185, Rome, Italy}
\affiliation[h]{Institute for Nuclear Research, 03028 Kyiv, Ukraine}
\affiliation[i]{IRFU, CEA, Universit\'e Paris-Saclay, F-91191 Gif-sur-Yvette, France}
\affiliation[j]{DISAT, Universit\`a dell'Insubria, I-22100 Como, Italy}
\affiliation[k]{INFN, Laboratori Nazionali del Gran Sasso, I-67100 Assergi (AQ), Italy}
\emailAdd{giuliani@csnsm.in2p3.fr}

\abstract{Neutrinoless double-beta decay is a key process in particle physics. Its experimental investigation is the only viable method that can establish the Majorana nature of neutrinos, providing at the same time a sensitive inclusive test of lepton number violation. CROSS (Cryogenic Rare-event Observatory with Surface Sensitivity) aims at developing and testing a new bolometric technology to be applied to future large-scale experiments searching for neutrinoless double-beta decay of the promising nuclei $^{100}$Mo and $^{130}$Te. The limiting factor in large-scale bolometric searches for this rare process is the background induced by surface radioactive contamination, as shown by the results of the CUORE experiment. The basic concept of CROSS consists of rejecting this challenging background component by pulse-shape discrimination, assisted by a proper coating of the faces of the crystal containing the isotope of interest and serving as energy absorber of the bolometric detector. In this paper, we demonstrate that ultra-pure superconductive Al films deposited on the crystal surfaces act successfully as pulse-shape modifiers, both with fast and slow phonon sensors. Rejection factors higher than 99.9\% of $\alpha$ surface radioactivity have been demonstrated in a series of prototypes based on crystals of Li$_2$MoO$_4$ and TeO$_2$. We have also shown that point-like energy depositions can be identified up to a distance of $\sim 1$~mm from the coated surface. The present program envisions an intermediate experiment to be installed underground in the Canfranc laboratory (Spain) in a CROSS-dedicated facility. This experiment, comprising $\sim 3\times 10^{25}$ nuclei of $^{100}$Mo, will be a general test of the CROSS technology as well as a worldwide competitive search for neutrinoless double-beta decay, with sensitivity to the effective Majorana mass down to 70~meV in the most favorable conditions.}

\begin{document} 
\maketitle
\flushbottom

\section{Introduction}
\label{intro}

Neutrinoless double-beta ($0\nu2\beta$) decay is a hypothetical rare nuclear transition, which plays a unique role in understanding fundamental neutrino properties and exploring lepton number violation (LNV)~\cite{Pas:2015a,Delloro:2016a,Vergados:2016a}. It consists in the transformation of an even-even nucleus into a lighter isobar containing two more protons and accompanied by the emission of two electrons and no other particles, with a change of the total lepton number by two units: $(A,Z)\to(A,Z+2)+2$e$^{-}$. The current most stringent half-life limits on $0\nu2\beta$ decay are of the order of $10^{25}$--$10^{26}$~y~\cite{Gando:2016a,Agostini:2018a,Alduino:2018a}. The standard process ($2\nu2\beta$ decay), which implies also the emission of two electron antineutrinos, is the rarest nuclear decay and has been observed in eleven nuclei with half-lives in the range 10$^{18}$--10$^{24}$~y~\cite{Barabash:2015a}. 

The detection of the neutrinoless channel would be a major discovery, and would represent the observation of a new phenomenon beyond the Standard Model (SM) of elementary particles, establishing that neutrino is a Majorana particle, rather than a Dirac one as all the other fermions: neutrino would be the only fermion to coincide with its own antimatter partner, a possibility left naturally open by its neutrality~\cite{Majorana:1937a}. In this framework, a new mechanism of mass generation, besides the Higgs mechanism, could be in place for neutrinos explaining naturally the smallness of ordinary neutrino masses (see for example Ref.~\cite{Bilenky:2011a}). In addition, matter-antimatter asymmetry in the Universe could be accounted for through CP violation in the neutrino sector~\cite{Fukugita:1986a}. 

It is important to remark however that, in a beyond-SM perspective, $0\nu2\beta$ decay is much more than a neutrino physics experiment. It is a powerful, inclusive test of LNV, which takes the form of a creation of electrons according to the process 2n~$\to$~2p~+~2e$^-$, implemented in the nuclear matter. LNV is as important as baryon number violation~\cite{DellOro:2017a} and naturally incorporated by beyond-SM theories. In that regard, the experimental search for $0\nu2\beta$ decay needs to be pursued with the highest possible sensitivity irrespectively of the related neutrino-physics scenario, as it is an essential element for a deep comprehension of the elementary constituents of matter and of fundamental interactions.

$0\nu2\beta$ decay can be induced by a plethora of LNV mechanisms. It is instructive to classify them in an effective-field-theory framework, in which the SM Lagrangian is extended by adding higher-order terms corresponding to increasing powers of a factor $1/\Lambda$, where $\Lambda$ is a cut-off high-energy scale, with respect to which the SM can be considered a low-energy approximation~\cite{Weinberg:1979a}. This approach allows us to consider all possible contributions to $0\nu2\beta$ decay and to include also interferences among different mechanisms (see for example Ref.~\cite{Cirigliano:2018a}). However, the lowest-order 1/$\Lambda$ term occupies a special place. It represents the so-called mass mechanism, a minimal extension of the SM naturally connected with neutrinos' being massive, as proved by flavor oscillations~\cite{Tanabashi:2018a}. In the mass mechanism, $0\nu2\beta$ decay is induced by the exchange of virtual light Majorana neutrinos. The rate of the process is proportional --- within an uncertainty due to the computation of the nuclear matrix elements --- to the square of the effective Majorana neutrino mass $m_{\beta\beta}$, related to the absolute neutrino mass scale and to the mass ordering~\cite{Pas:2015a,Delloro:2016a,Vergados:2016a}. 

Present limits on $m_{\beta\beta}$ from $0\nu2\beta$ decay are in the range $\sim$60--600~meV~\cite{Gando:2016a,Agostini:2018a,Alduino:2018a,Auger:2012a,Aalseth:2018a,Arnold:2015a}, assuming that the axial charge $g_A$ is not quenched --- the $0\nu2\beta$ decay rate is proportional to $g_A^4$ --- and close to the free nucleon value of ~1.27 (the most common approach in the literature). The possible quenching of $g_A$, which could reduce even by a factor $\sim 4$ the sensitivity to $m_{\beta\beta}$, is a controversial open issue~\cite{Engel:2017a,Suhonen:2017a}. In particular, it is not clear if the quenching observed in single $\beta$ decay and $2\nu2\beta$ decay can be naively extended to $0\nu2\beta$ decay, which involves a much larger momentum transfer than in the ordinary processes. Furthermore, this quenching --- even if present in the mass mechanism --- could have no impact on other LNV mechanisms. A spread of a factor 2--3 in the limit on $m_{\beta\beta}$ is inherently connected to our approximate knowledge of the nuclear physics involved in the $0\nu2\beta$ transition~\cite{Engel:2017a}. None of the current experiments has the potential to explore fully the inverted-ordering region of the neutrino mass pattern --- corresponding to the range 15--50 meV for $m_{\beta\beta}$ --- even in case of no $g_A$ quenching~\cite{Giuliani:2018a,Barabash:2019a}. In addition, we should be prepared to the fact that the ordering could be normal, as preliminary suggested by recent results of neutrino oscillation experiments with accelerator sources~\cite{Acero:2018a} and by cosmology~\cite{Palanque_Delabrouille:2015a}. In that regard, we remark that exploring fully the inverted-ordering band implies probing simultaneously $\sim$~50\% of the normal-ordering region, according to a Bayesian analysis based on our current experimental knowledge of the constraints on the neutrino mass and oscillation parameters~\cite{Agostini:2017a}. 

In this challenging context, it is crucial to explore new experimental ideas that could extend the sensitivity of future searches. In this article, we present an innovative bolometric technology to be tested in a demonstrator named CROSS (Cryogenic Rare-event Observatory with Surface Sensitivity). CROSS will use lithium molybdate (Li$_2$MoO$_4$) and tellurium oxide (TeO$_2$) crystals, operated as bolometric detectors, with the addition of an Al-film coating in order to induce signal-shape modification of nuclear events occurring close to the detector surface. This will allow us to get rid of the currently dominant background component in bolometric experiments. 

This paper is organized as follows. In Section~\ref{sec:concepts} we will describe the concepts on which CROSS is based: preliminary discussion on the background in bolometric detectors, consequent choice of the isotopes and of the materials, and illustration of the methods adopted to achieve sensitivity to surface events. In the subsequent Section~\ref{sec:results}, we will describe the preliminary encouraging results obtained by coating the crystals with Al films. We will discuss then --- in Section~\ref{sec:prospect} --- the structure and the sensitivity of the planned CROSS demonstrator and the prospects for a future ton-scale experiment. 

\section{Concepts of the CROSS technology}
\label{sec:concepts}

In order to observe $0\nu2\beta$ decay, experimentalists aim at the detection of the two emitted electrons, which share the total transition energy (the so-called $Q$-value of the process)~\cite{Pas:2015a,Delloro:2016a,Vergados:2016a,Giuliani:2012a,Cremonesi:2013a}. The signature is a peak at the $Q$-value in the sum-energy spectrum of the two electrons. Bolometers, the instruments chosen for CROSS, are among the most powerful nuclear detectors for the conduction of sensitive $0\nu2\beta$ decay searches in the calorimetric approach~\cite{Fiorini:1984}. They can provide high sensitive mass (via large detector arrays), high detection efficiency, high energy resolution and extremely low background thanks to potentially high material radiopurity and methods to reject parasitic events~\cite{Poda:2017a}. A $Q$-value as high as possible is desirable since this places the signal in a lower background region and increases strongly the decay probability. This feature can be met by bolometers as most of the high $Q$-value candidates ($^{48}$Ca, $^{76}$Ge, $^{82}$Se, $^{96}$Zr, $^{100}$Mo, $^{116}$Cd, $^{124}$Sn, $^{130}$Te) can be studied with this technique.
 
A bolometer consists of a single dielectric crystal --- the active part of the detector --- coupled to a phonon sensor. The signal, collected at very low temperatures ($< 20$~mK for large bolometers, with masses in the 0.1--1 kg range), consists of a phonon pulse registered by the sensor. In CROSS, we will use two types of phonon sensors: (i) a neutron transmutation doped (NTD) Ge thermistor~\cite{Haller:1984a} (following the scheme adopted in Cuoricino and CUORE~\cite{Alduino:2018a}), which is mainly sensitive to thermal phonons and can be considered with a good approximation as a thermometer, and (ii) a NbSi thin film~\cite{Dumoulin:1993a,Crauste:2011}, which is faster and exhibits a significant sensitivity to athermal phonons.\footnote{The energy spectrum of phonons excited by a particle event is highly non-thermal immediately after the particle interaction, but it tends to relax to a thermal-equilibrium distribution at long times. A deeper discussion can be found in Section~\ref{sec:surface-sensitivity}.}

\subsection{Background in bolometric detectors}
\label{sec:background}

Environmental $\gamma$'s represent the main source of background for most of the present experiments and arise mainly from natural contamination in $^{238}$U and $^{232}$Th and their daughters. The highest-energy intense $\gamma$ line in natural radioactivity is the 2615 keV line of $^{208}$Tl, belonging to the $^{232}$Th decay chain. Therefore, a detector embedding a candidate with a $Q$-value above 2615 keV represents an optimal choice. However, the energy region above $\sim$~2.5 MeV is dominated by events due to surface radioactive contamination, especially $\alpha$ particles (as shown by the results of the bolometric $0\nu2\beta$-decay search CUORE~\cite{Alduino:2018a} and its predecessors Cuoricino and CUORE-0), but also $\beta$ particles emitted at the surface can be the limiting factor in future searches.  
An extensively studied and adopted method to reject surface $\alpha$ events in bolometers exploits the scintillation~\cite{Pirro:2006a} or Cherenkov radiation~\cite{TabarellideFatis:2009a} emitted by detector materials~\cite{Artusa:2014a,Poda:2017a}. Since the $\alpha$ light yield is appreciably different from (in general inferior to) the $\beta$ light yield at equal deposited energy, while the bolometric thermal response is substantially equivalent, the simultaneous detection of light and heat, and the comparison of the respective signal amplitudes, represent a powerful tool to reject $\alpha$ background (in particular from surface) through $\alpha$/$\beta$ discrimination. However, this method has two drawbacks: (i) an additional device --- a light detector consisting of an optical bolometer --- is required, and (ii) surface $\beta$ events are not rejected. A key advantage of the CROSS technology is to provide the bolometers with surface sensitivity, enabling in principle the mitigation of both the $\alpha$ and $\beta$ background components without supplementary devices. In practice, we demonstrate in this paper that the surface $\alpha$ component can be fully rejected with the CROSS method, while the identification of $\beta$ particles releasing their energy at the detector surface requires additional R\&D work.

We can already evaluate the effect of $\alpha$ rejection on the background in a real large-scale bolometric set-up like the CUORE infrastructure. CUORE, comprising almost 1000 individual bolometers, is the largest-ever bolometric experiment~\cite{Alduino:2018a}. It collects data in the Gran Sasso underground laboratory (Italy) and investigates the $0\nu2\beta$ decay of the nuclide $^{130}$Te. After about 90 kg$\times$y exposure, a preliminary background model~\cite{Benato:2019a} can be used to show that, in the current CUORE infrastructure, the expected background index would be $\sim 10^{-4}$ counts/(keV kg y) in the region of interest (ROI) of $^{100}$Mo at $\sim 3$~MeV after surface $\alpha$ background rejection. Conversely, a background level an order of magnitude higher is expected in the ROI of $^{130}$Te at $\sim 2.5$~MeV, because of a residual contamination of $^{232}$Th in the CUORE cryostat, whose location is still under investigation. This contamination contributes to the background almost only for $^{130}$Te, due to the lower $Q$-value of this isotope. 

\subsection{Isotope and compound choice in CROSS}
\label{sec:isotope}

The choices of the candidates and of the compounds for CROSS are based on years' development of the bolometric search for $0\nu2\beta$ decay (projects MIBETA~\cite{Arnaboldi:2003a}, Cuoricino~\cite{Arnaboldi:2008a}, CUORE~\cite{Alduino:2018a}, LUCIFER/CUPID-0~\cite{Azzolini:2018a}, LUMINEU/CUPID-Mo~\cite{Armengaud:2017a,Armengaud:2019a}). A critical analysis of the copious results achieved so far on more than 10 compounds indicates that the best choices are TeO$_2$~\cite{Alduino:2018a,Alessandria:2012a} (containing the candidate $^{130}$Te) and Li$_2$MoO$_4$~\cite{Armengaud:2017a,Poda:2017b,Armengaud:2019a} (containing the candidate $^{100}$Mo). 

TeO$_2$ has been extensively and successfully studied as it is the material chosen for CUORE, as discussed above. Detectors based on TeO$_2$ exhibit excellent energy resolution ($\Delta_{FWHM} \sim 5$~keV in the ROI around the $^{130}$Te $Q$-value at 2527~keV) and very high internal purity~\cite{Alessandria:2012a} (< 1 $\mu$Bq/kg for $^{238}$U and $^{232}$Th and their daughters). Preliminary results on TeO$_2$ crystals grown with tellurium enriched in $^{130}$Te at 92\% level reproduce the excellent performance of the devices with natural isotopic composition~\cite{Artusa:2017a}. A drawback of $^{130}$Te is that its $Q$-value is just below the 2615 keV line of $^{208}$Tl. Therefore, the control of the $^{232}$Th contamination is a major issue and a possible show-stopper, as the aforementioned results of the CUORE background model indicate clearly. Another drawback of $^{130}$Te is related to the compound TeO$_2$, which features a very low scintillation yield~\cite{Coron:2004a,Berge:2018a}. Consequently, an optical bolometer needs to detect the tiny Cherenkov radiation emitted by $\beta$ particles~\cite{TabarellideFatis:2009a} in order to exclude the $\alpha$ component in a future upgrade of TeO$_2$ bolometers. This requires an intense R\&D that is far to be concluded at the moment~\cite{Poda:2017a}.  

As for Li$_2$MoO$_4$, the main advantage is that $^{100}$Mo has a $Q$-value of 3034 keV, above the highest-energy line of $^{208}$Tl. The merits of this compound were evidenced recently ~\cite{Bekker:2016a} and extensively tested in the scintillating-bolometer LUMINEU and CUPID-Mo projects with large-volume, high-quality crystals containing molybdenum enriched in $^{100}$Mo at 97--99\% level~\cite{Armengaud:2017a,Poda:2017b,Armengaud:2019a}. The energy resolution achieved on prototypes is excellent ($\Delta E_{FWHM} \sim 5$~keV in the ROI) and the radiopurity very high ($< 3 \ \mu$Bq/kg for $^{238}$U and $^{232}$Th and their daughters). The crystallization process is simple and the irrecoverable losses of the costly $^{100}$Mo are negligible ($< 4$\%), implying low-cost crystal production and high production rate. (Ref.~\cite{Berge:2014a} describes the protocol for ZnMoO$_4$, which can be extended smoothly to Li$_2$MoO$_4$~\cite{Grigorieva:2017a}.) A drawback of $^{100}$Mo is the relatively fast $2\nu2\beta$ process in this isotope (T$_{1/2}$~$\sim 7 \times 10^{18}$~y~\cite{Barabash:2015a}). This induces a background through random coincidences of the ordinary $2\nu2\beta$ events~\cite{Chernyak:2012a}.

Both $^{130}$Te and $^{100}$Mo do not pose problems for massive enrichment at viable costs~\cite{Giuliani:2012a}. We stress that the natural isotopic abundance of $^{130}$Te is very high, of the order of 34\% (about 3 times higher than that of $^{100}$Mo)~\cite{Meija:2012a}.  

CROSS intends to study both options, in order to guide the selection for future bolometric experiments or to provide the elements for a twofold large scale search~\cite{Giuliani:2018b,CUPID:2019a}. 
The CROSS solution against surface radioactivity, which does not imply light detection (see Section~\ref{sec:surface-sensitivity}), is specially convenient for the case of TeO$_2$, due to the difficulties to detect Cherenkov light. Of course, the CROSS approach cannot control the $\gamma$ background, but the uniquely high natural isotopic abundance of  $^{130}$Te suggests to keep the CROSS technology ready for this compound with a view to large-scale future experiments that could be hosted in an extremely radiopure environment with low $\gamma$ emission. The CROSS methods can help control the $2\nu2\beta$-induced background in the $^{100}$Mo case by adopting NbSi thin films as temperature sensors, as they can guarantee fast pulse rise-times --- of the order of the millisecond --- even in large crystals. However, there are good prospects to reject this background component at the desired level even with a readout based on NTD Ge thermistors~\cite{Chernyak:2014a}.

\subsection{Discrimination of surface radioactivity in CROSS}
\label{sec:surface-sensitivity}

CROSS aims at developing a technology capable of identifying and rejecting $\alpha$ surface events with an efficiency higher than 99.9\%, without a significant change of the acceptance of $0\nu2\beta$ decay signals. As we will show in Section~\ref{sec:rejalpha}, this result has already been achieved in low-mass prototypes both for Li$_2$MoO$_4$ and TeO$_2$. The physics of the rejection mechanism could allow us to reject in principles also events occurring down to a depth of about 1~mm from the crystal surface, making the control of $\beta$ surface background possible as well. This result has not been achieved yet, although there are hints that the surface sensitivity extends to spot-like energy depositions (within $\sim 100$~$\mu$m) occurring down to $\sim 1.4$~mm from the surface in Li$_2$MoO$_4$ crystals (Section~\ref{sec:neutron}). 

The discrimination of surface events will be obtained by depositing thin films, with a thickness in the few-$\mu$m range, on the surfaces of the Li$_2$MoO$_4$ and TeO$_2$ crystals. The material of the film will be superconductive aluminum (Al critical temperature is $\sim 1.2$~K~\cite{Cochran:1958a}).  The proof of concept of CROSS surface sensitivity was preliminarily achieved in a test performed a few years ago at CSNSM (Orsay, France), using a fast thermal sensor ($\sim 1$~ms time response) coupled to a TeO$_2$ crystal~\cite{Nones:2012a}. 

\begin{figure}[tbp]
\centering 
\includegraphics[width=0.7\textwidth]{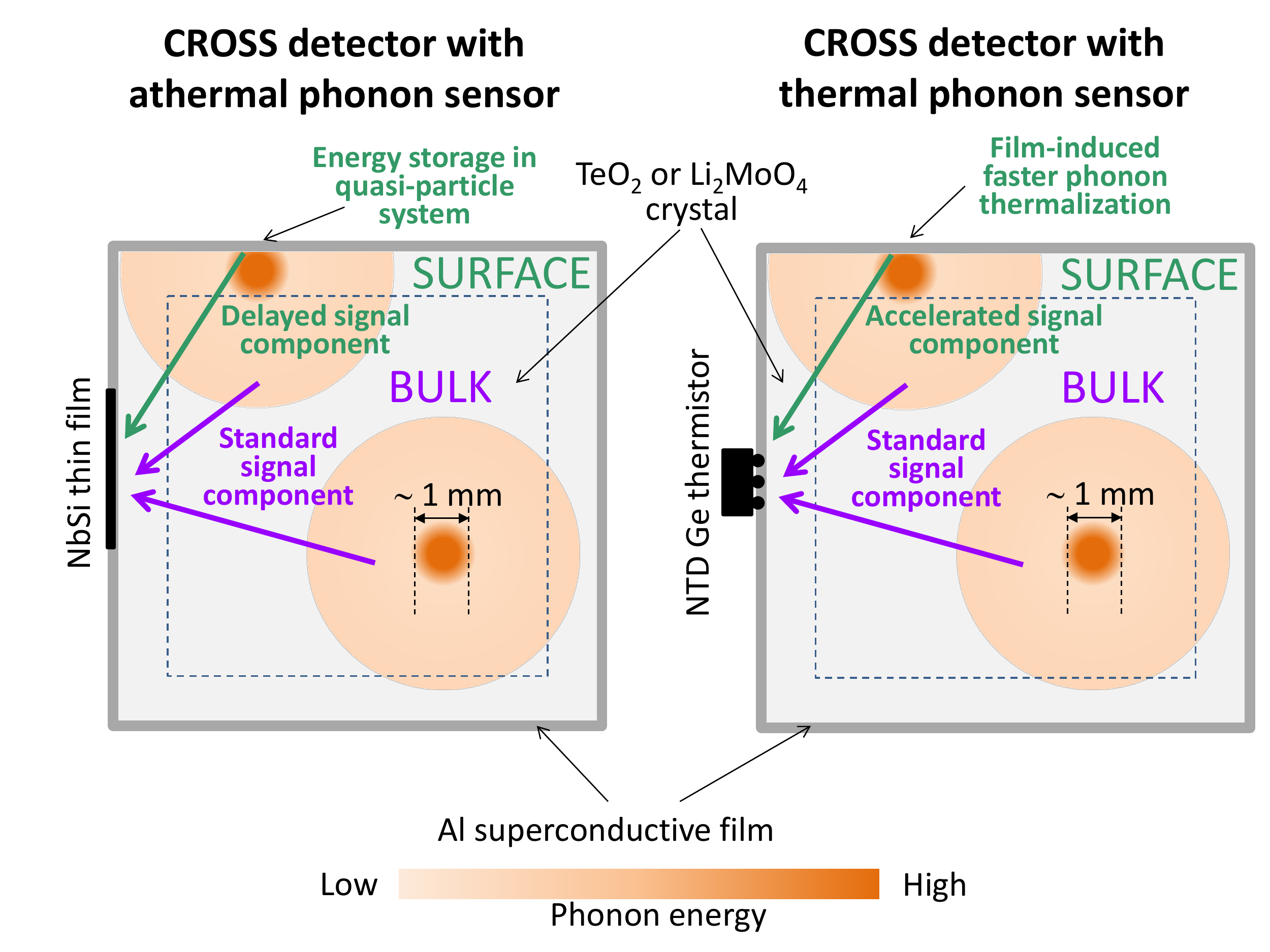}
\caption{\label{fig:phonon-scheme} Qualitative scheme of the mechanism of surface-event identification in CROSS. The time response of the phonon sensors depends on the distance of the particle energy deposit from the superconductive Al film. If the $\sim$~1-mm-size phonon hot spot reaches the film, an athermal-phonon sensor will provide a delayed signal with respect to a bulk event (left panel) due to energy trapping in the form of quasi-particles. Conversely, a thermal-phonon sensor will provide a faster signal with respect to a bulk event (right panel) as the film accelerates the phonon thermalization process. Signals from athermal-phonon sensors are anyway globally faster than those from thermal-phonon ones.}
\end{figure}

The rationale of the discrimination capability is based on the properties of non-equili-brium phonons evolving from a hot spot (see Fig.~\ref{fig:phonon-scheme}, left panel), generated by a particle interaction in a non-metallic material~(\cite{Tamura:1997a} and references therein). The athermal phonons produced by a nuclear event have initially (after a few $\mu$s) ``high'' energies ($> 30$~K) in a cloud of $\sim 1$~mm size around the event location. These energetic prompt phonons will reach the superconductive film and break efficiently Cooper pairs in it when the event is located within $\sim 1$~mm from the surface. (We stress that this is the typical range of a $\sim$~MeV  $\beta$~particle impinging on the detector and emitted by a surface radioactive impurity.) This effect is less important for energy depositions occurring at a longer distance from the surface (bulk events) since in this case the particle-generated phonons reach the film with a significantly lower average energy, due to the quasi-diffusive mode of phonon propagation, which implies spontaneous fast phonon decay when the phonon energy is high. In fact, the phonon decay rate is proportional to the fifth power of the phonon energy~\cite{Orbach:1964a}. Due to the long recombination time of quasiparticles at temperature well below the transition temperature of the superconductor, a significant fraction of the particle energy will be trapped in the superconductive layer in the form of quasiparticles for a time that can extend up to the millisecond range in ultrapure aluminum. The subsequent recombination of quasiparticles to $\sim 1.2$~K phonons will add a delayed component to the phonon signal read by the sensor on the crystal~\cite{Schnagel:2000a}, less important in bulk events. Therefore, just by studying the signal shape and in particular the rise-time, it will be possible to discriminate bulk versus surface events, as the latter should be slower with a rise-time difference in the millisecond range. We stress that this mechanism works for sensors that are sensitive to athermal phonons, like thin films directly deposited on the crystal surface and capable of intercepting efficiently the delayed component, which could represent a small fraction of the total deposited energy. 

As reported in Section~\ref{sec:results}, we initially attempted pulse shape discrimination (PSD) using NTD Ge thermistors, as these sensors, even if mainly measuring the crystal temperature, may exhibit sensitivity to athermal phonons. This choice was dictated by the simplicity of this option, which would allow us to keep the present CUORE configuration in terms of readout. We have discovered that surface sensitivity can in fact be provided by NTD Ge thermistors but with a different mechanism (see Fig.~\ref{fig:phonon-scheme}, right panel), as shown by the fact that the leading edge of surface-event signals is indeed faster --- and not slower as predicted --- than bulk-event signals in the NTD-Ge-thermistor case. The test described in Section~\ref{sec:NbSiNTD} provides a viable interpretation of this mechanism.

\section{Preliminary results on the identification of surface events} 
\label{sec:results}

\subsection{Experimental set-up and prototype construction}
\label{sec:set-up}

\begin{figure}[tbp]
\centering 
\includegraphics[width=0.7\textwidth]{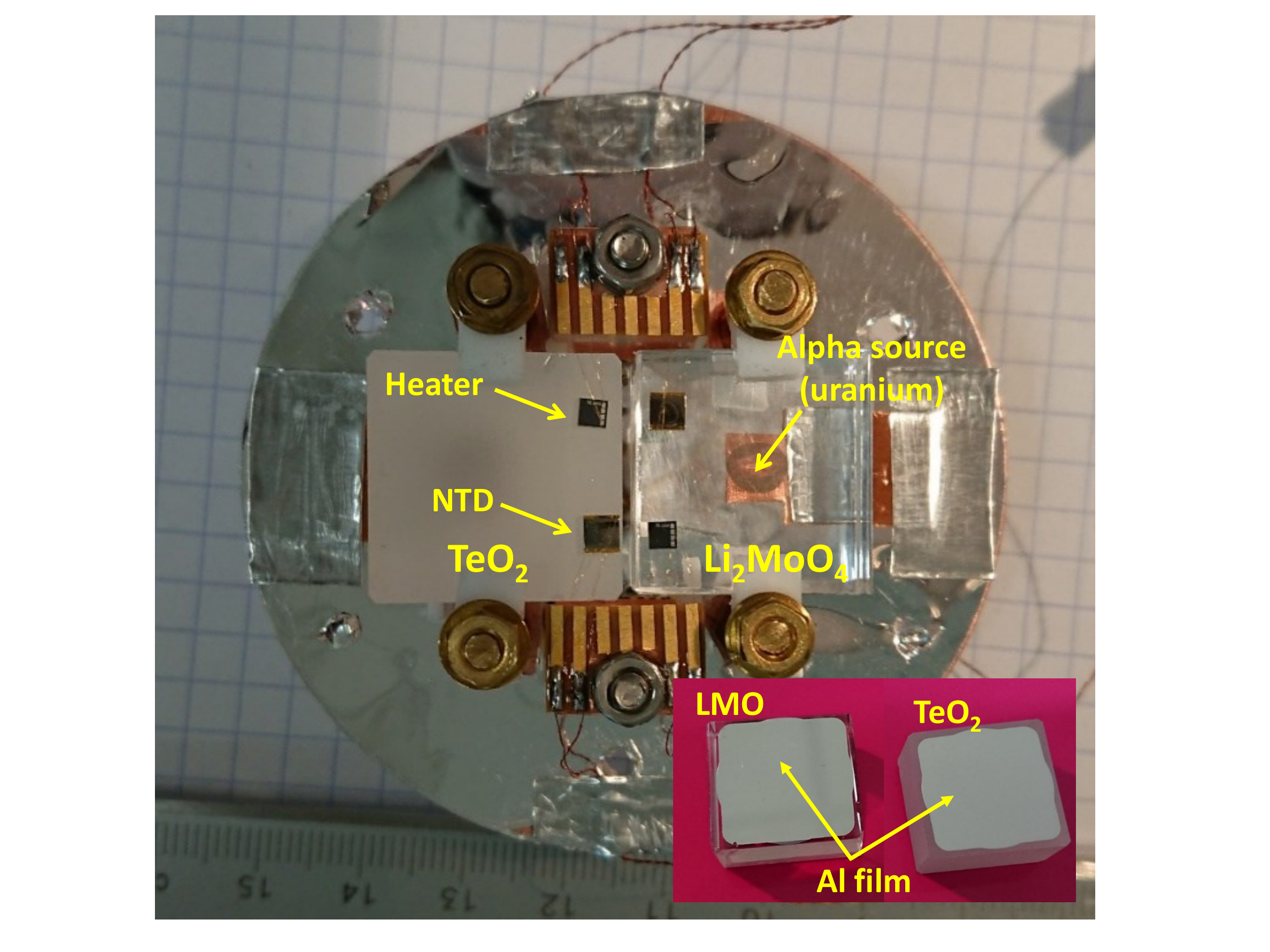}
\caption{\label{fig:photo} Photograph of the set-up used for the CROSS-prototype tests. The Li$_2$MoO$_4$ and TeO$_2$ crystals, with a size of 20$\times$20$\times$10 mm, are visible. The NTD Ge thermistors and the heaters are glued on the upper crystal surface. The uranium sources are placed below the crystals (the one irradiating Li$_2$MoO$_4$ is visible in transparency). No Al film is deposited in this case. The copper disk which supports the crystals is covered with a reflective foil. Two same-size crystals can be assembled symmetrically below the copper disk. A light detector (not shown here), consisting of a $\oslash$44$\times$0.175-mm Ge wafer coated by a SiO thin layer to reduce the light reflection, can face two crystals simultaneously at both sides of the copper disk. The inset at the bottom right corner shows Li$_2$MoO$_4$ and TeO$_2$ crystals with a 20$\times$20 mm side coated by a 10~$\mu$m-thick Al film.}
\end{figure}

We have designed and fabricated a set-up (depicted in Fig.~\ref{fig:photo}) to test simultaneously up to four crystals (Li$_2$MoO$_4$ or TeO$_2$) with 20$\times$20$\times$10~mm or 20$\times$20$\times$5~mm size. For the moment, we have operated up to three detectors in each run. Both Li$_2$MoO$_4$ and TeO$_2$ crystals are grown with materials of natural isotopic composition. The samples of Li$_2$MoO$_4$ are cut from a large cylindrical crystal grown with the low-thermal gradient Czochralski technique at Nikolaev Institute of Inorganic Chemistry (Novosibirsk, Russia)~\cite{Grigorieva:2017a} and similar to the samples employed in the LUMINEU and CUPID-Mo projects. The samples of TeO$_2$ are produced by the company SICCAS (Shanghai, China) following the same protocol employed for the crystals of the CUORE experiment~\cite{Alessandria:2012a}.

Even if CROSS intends to perform background rejection without optical bolometers, a light detector has often been used in the tests described below in order to discriminate $\alpha$ particles and neutron interactions on $^{6}$Li (see Section~\ref{sec:neutron}) and help in the result interpretation. The light absorber of the optical bolometer consists of a high-purity germanium wafer ($\oslash$44$\times$0.175 mm). The light detector is often used in the Neganov-Trofimov-Luke mode with the purpose to enhance the device sensitivity~\cite{Artusa:2017a,Berge:2018a,Novati:2019a}. This is convenient considering the feeble light emitted by the TeO$_2$ crystal and the complicated geometry of the assembly that does not ensure optimal light collection. The light detector helps also to veto cosmic muon events. 

The crystals and the Ge wafer are fixed at the copper heat sink by means of PTFE pieces and brass / copper screws. The inner surfaces of the detector holder are covered as much as possible by a reflecting foil (Vikuiti$^{\rm TM}$ Enhanced Specular Reflector Film) in order to improve light collection.

With the exception of the test described in Section~\ref{sec:NbSiNTD}, the phonon sensors used for the Li$_2$MoO$_4$ and TeO$_2$ crystals and for the Ge wafer are exclusively NTD Ge thermistors. These elements have a mass of about 50~mg when used for the large crystals and about a factor 10 lower for the light-detector wafers. Their resistance is $\sim 1$ M$\Omega$ at 20~mK. They are glued at the crystal surface by using six epoxy spots and a 25-$\mu$m-thick Mylar spacer, which is removed after the gluing procedure. In addition, each crystal is provided with a resistive element (heater) capable of delivering periodically a fixed amount of thermal energy in order to control and stabilize the response of the bolometers~\cite{Alessandrello:1998a}. 

In order to induce sensitivity to surface events, an aluminum layer is evaporated under vacuum --- the residual pressure is typically $5 \times 10^{-7}$~mbar --- on a crystal side, by means of an electron gun located at CSNSM. Before the deposition, the surface of the crystal is bombarded in situ by 90-eV Ar ions that remove a layer of about 1~nm. In the case of 10-$\mu$m-thick films, a 25-nm-thick SiO sub-layer is deposited after the Ar bombardment in order to improve the Al-film adhesion. The Al deposition rate is of 4~nm/s and the layer thickness of 1 or 10 $\mu$m depending on the test. The distance between the source and the sample is about 45~cm and the control of the thickness is done by looking at the evaporation duration with a periodic check of the rate through a piezoelectric quartz. The Li$_2$MoO$_4$ crystals are polished before Al deposition to avoid effects due to the weak hygroscopicity of Li$_2$MoO$_4$, while the TeO$_2$ surfaces underwent only a rough grinding to imitate the opaque diffusive surfaces that characterize the CUORE crystals.

Uranium sources, obtained by desiccating a drop of uranium acid solution on a copper foil, are placed in front of the Al coated surfaces, at about 0.6 mm distance. They provide two main $\alpha$ lines at $\sim 4.2$ and $\sim 4.7$~MeV ($^{238}$U and $^{234}$U respectively) but also $\beta$ particles from the isotope $^{234m}$Pa, with a spectrum extending up to $\sim 2.27$~MeV. The $^{234m}$Pa $\beta$ rate is very close to the $\alpha$ rate of $^{238}$U. Another $\alpha$ source has been obtained by implanting $^{218}$Po atoms (produced by $^{222}$Rn decay) on a copper tape, which had been put in a radon-rich atmosphere by emanation from a uranium mineral. The $^{218}$Po and its daughters decay quickly to $^{210}$Pb, which has a half-life of 22.3 years. Its decay leads to $^{210}$Po that provides the desired $\sim 5.3$-MeV $\alpha$ line. The polonium sources are usually placed in front of an uncoated side. They are expected therefore to generate events with the same pulse shape as bulk events. The two event classes (close or far from the Al film) can be easily distinguished also by their energy. In the following, for brevity, we will often define ``surface events'' the energy depositions occurring at the vicinity of the Al film and identifiable by pulse shape.

All the tests here discussed were performed in the cryogenic laboratory of CSNSM in a dry high-power dilution refrigerator with a 4~K stage cooled by a pulse-tube~\cite{Mancuso:2014a}. We made most of the measurements with the sample-holder temperature stabilized at 15.5 mK and 22 mK. Higher temperatures (30--40 mK) were used for the test reported in Section~\ref{sec:NbSiNTD} to match at best the resistance-temperature curve of the NbSi film, operated in the Anderson-insulator mode~\cite{Dumoulin:1993a}. The sample holder is mechanically decoupled from the mixing chamber by four springs to reduce the acoustic noise caused by the cryostat vibrations. The outer vacuum chamber of the refrigerator is surrounded by a passive shield made of low radioactivity lead (10~cm minimum thickness) to suppress the environmental $\gamma$ background. The shield mitigates the pile-up problem typical for above-ground measurements with macro-bolometers. For the same reason, we have used relatively small samples to reduce the counting rate of the environmental $\gamma$ background.

A room-temperature electronics based on DC-coupled low-noise voltage-sensitive amplifiers is used in the experiment~\cite{Arnaboldi:2002a}. The thermistors are biased via a constant current injected through two room-temperature load resistors in series, with a total resistance of typically 200~M$\Omega$.  The amplifier output data are filtered by a Bessel low-pass filter with a high frequency cut-off at 675~Hz and acquired by a 16~bit ADC with 10~kHz sampling frequency.

The pulse amplitudes of the signals are extracted off line using an optimal-filter analysis~\cite{Gatti:1986a}. This requires the use of a template for the noise power spectrum and for the average pulse. Since in the tests here reported pulses have often slightly different pulse shapes depending on the event type (typically, surface vs. bulk), it may happen --- when the pulse shape of the analyzed pulse differs from that of the template --- that the extracted amplitude is not a fully faithful estimator of the unfiltered-pulse height. This discrepancy does not affect critically the results reported below.   

We have performed several runs in order to study and optimize surface sensitivity. We will report about the results of four coated detectors, described in Table~\ref{tab:summary}. We performed a preliminary run with bare detectors as well, using as much as possible the same configuration as that adopted in tests with Al films, in order to establish a benchmark in terms of pulse shapes, detector sensitivity and intrinsic energy resolution.   

\begin{table}[tbp]
\centering
\begin{tabular}{|c|cccccc|}
\hline
\ & \ & Crystal & Crystal & Al Film & Al Film & Sensor  \\
Detector & Compound & size & mass & thickness & area & type  \\
\ & \ & $[$mm$]$ & $[$g$]$ & $[\mu$m$]$ & $[$mm$]$ & \  \\
\hline 
1 & Li$_2$MoO$_4$ & 20$\times$20$\times$10 & 12 & 10 & 20$\times$20 & NTD~Ge  \\
2 & TeO$_2$ & 20$\times$20$\times$10 & 25 & 10 & 20$\times$20 & NTD~Ge  \\
3 & TeO$_2$ & 20$\times$20$\times$10 & 25 & 1 & 20$\times$20 & NTD~Ge  \\
4 & TeO$_2$ & 20$\times$20$\times$5 & 12.5 & 10 & 20$\times$5 & NTD~Ge+NbSi  \\
\hline
\end{tabular}
\caption{\label{tab:summary} Main characteristics of the four detectors described in the text. An uranium $\alpha$ source irradiates the film in all the detectors. A polonium (uranium) $\alpha$ source irradiates a bare side of the crystal in detector 2 (4).}
\end{table}


\subsection{Rejection of surface events by Al-film coating}
\label{sec:rejalpha}

We have tested PSD with Li$_2$MoO$_4$ and TeO$_2$ crystals with one $20 \times 20$~mm side coated with an Al film. The coated surface represents $1/4$ of the total crystal surface. In the first test, the film thickness was 10~$\mu$m for both crystals (Detectors 1 and 2 in Table~\ref{tab:summary}). Subsequently, we reduced the thickness down to 1~$\mu$m for the TeO$_2$ crystal (Detector 3 in Table~\ref{tab:summary}). A clear surface-$\alpha$-event rejection was demonstrated in all cases.

\begin{figure}[tbp]
\centering 
\includegraphics[width=0.49\textwidth]{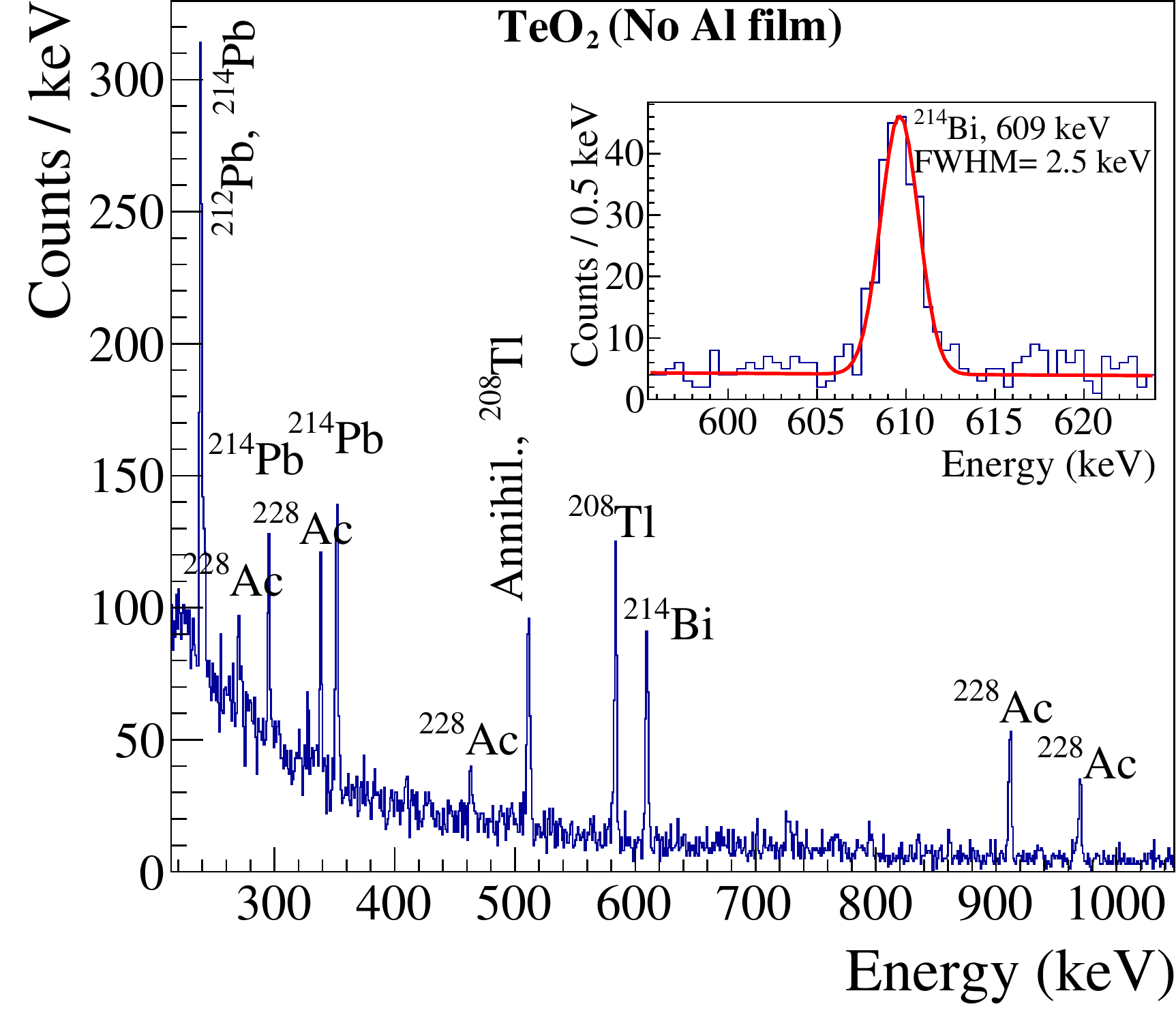}
\includegraphics[width=0.49\textwidth]{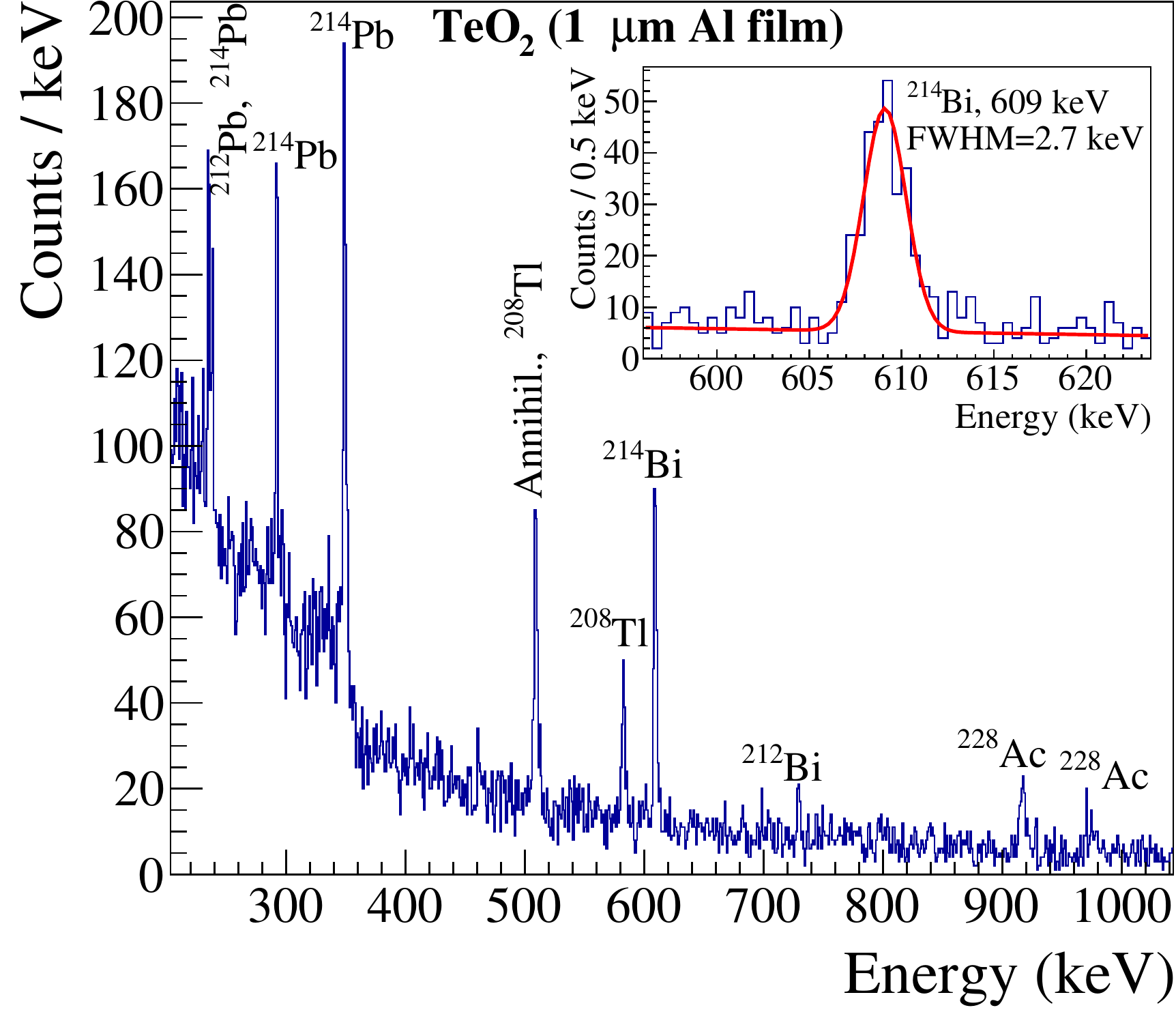} 
\caption{\label{fig:comparison} Left panel: $\gamma$ spectrum collected with a bare-crystal TeO$_2$ bolometer. Peaks generated by the environmental radioactivity and by a $^{232}$Th calibration source are appreciable. Right panel: $\gamma$ spectrum collected in similar environmental conditions with an identical TeO$_2$ bolometer but with one 20$\times$20 mm crystal side coated by a 1-$\mu$m-thick Al film. In the insets, the 609~keV $\gamma$ peaks of $^{214}$Bi are shown to compare the energy resolutions.} 
\end{figure}

By comparison with the blank test, we have first shown that there is a negligible influence of the Al films --- independently of their thickness --- on the detector behavior. In Fig.~\ref{fig:comparison} we show two spectra taken with the TeO$_2$ detectors during a $^{232}$Th calibration, without and with 1-$\mu$m-thick Al film. The energy resolution is unchanged, apart from variations at the $\sim 10$~\% level that are physiological in this technology from run to run, especially if bolometers are disassembled and reassembled. The energy resolution is about 2--3 keV FWHM for both configurations at the $^{214}$Bi 609~keV $\gamma$ peak, induced by the environmental radioactivity. At higher energies, we have peaks affected by low statistics because of the reduced size of the detectors and sometimes enlarged by non-fully efficient stabilization. In Table~\ref{tab:behavior}, we report the comparison of the sensitivities (expressed as pulse-amplitude/deposited-energy [nV/keV]) and of the rise-times $t_R$ (from 10\% to 90\% of the pulse maximum amplitude) and decay-times $t_D$ (from 90\% to 30\% of the pulse maximum amplitude) for $\gamma$ events in the measurements performed at 22~mK, for which we were able to reasonably reproduce the operation points in the two runs. The sensitivities and the decay-times are similar, once again within the typical variations due to the non-complete reproducibility of the experimental conditions. The shortening of the rise-time in the case of Li$_2$MoO$_4$ after film deposition is not fully understood, but it may be related to the improved-thermalization mechanism described in Section~\ref{sec:NbSiNTD}.

\begin{table}[tbp]
\centering
\begin{tabular}{|c|cccc|}
\hline
Detector & Al & Sensitivity & $t_R$ & $t_D$ \\
crystal  & film & $[$nV/keV$]$ & $[$ms$]$ & $[$ms$]$ \\
\hline 
Li$_2$MoO$_4$ & NO & 57 & 8.4 & 34 \\
Li$_2$MoO$_4$ & $10 \ \mu$m & 53 & 3.1 & 24 \\
\hline
TeO$_2$ & NO & 47 & 7.9 & 69 \\
TeO$_2$ & $10 \ \mu$m & 43 & 7.2 & 59 \\
\hline
\end{tabular}
\caption{\label{tab:behavior} Sensitivities and pulse-shape parameters of detectors without and with 10-$\mu$m-thick Al film. The crystal size is 20$\times$20$\times$10 mm and the measurement is performed with the detector-holder temperature stabilized at 22 mK.}
\end{table}

\begin{figure}[tbp]
\centering 
\includegraphics[width=0.49\textwidth]{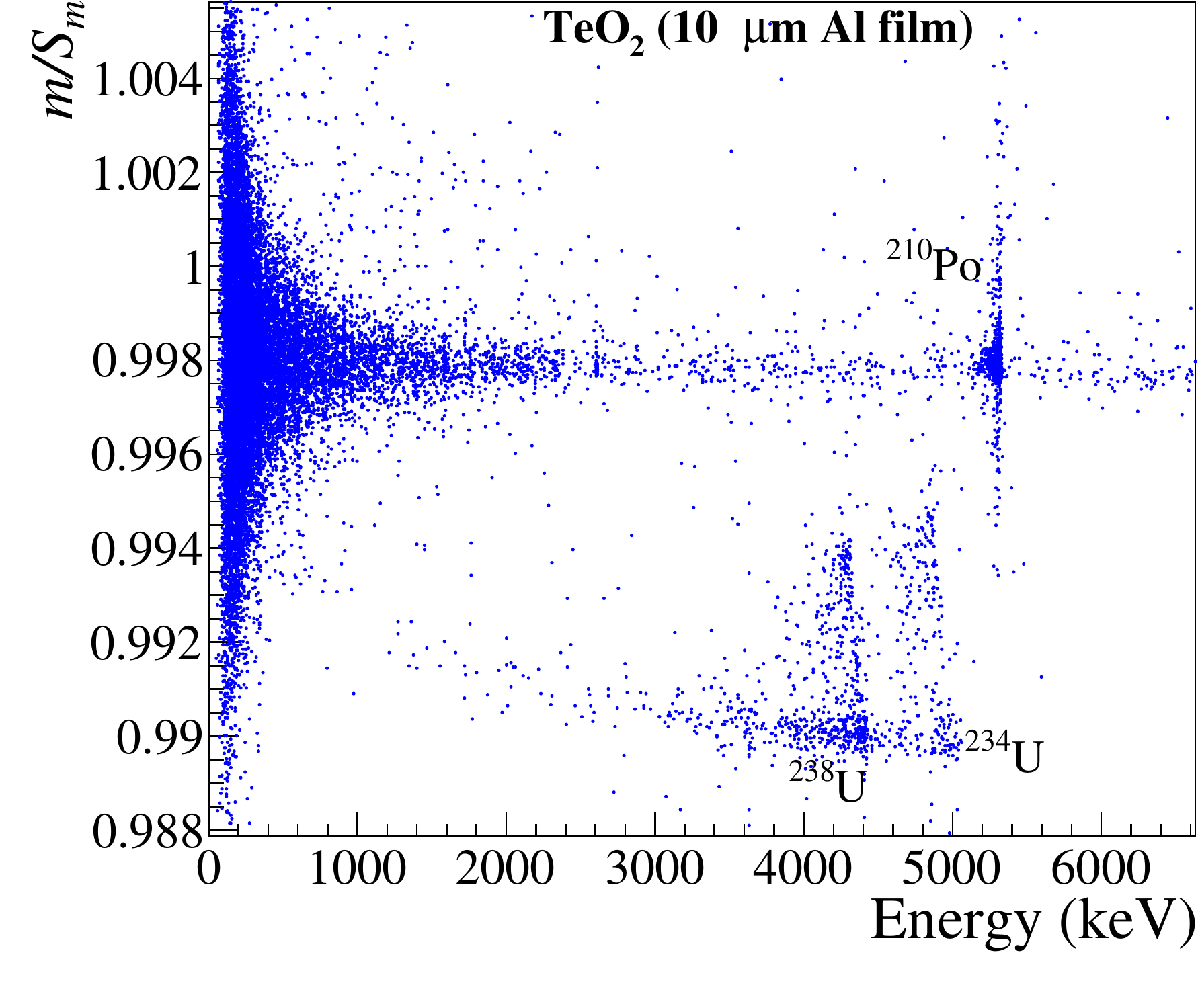}
\includegraphics[width=0.49\textwidth]{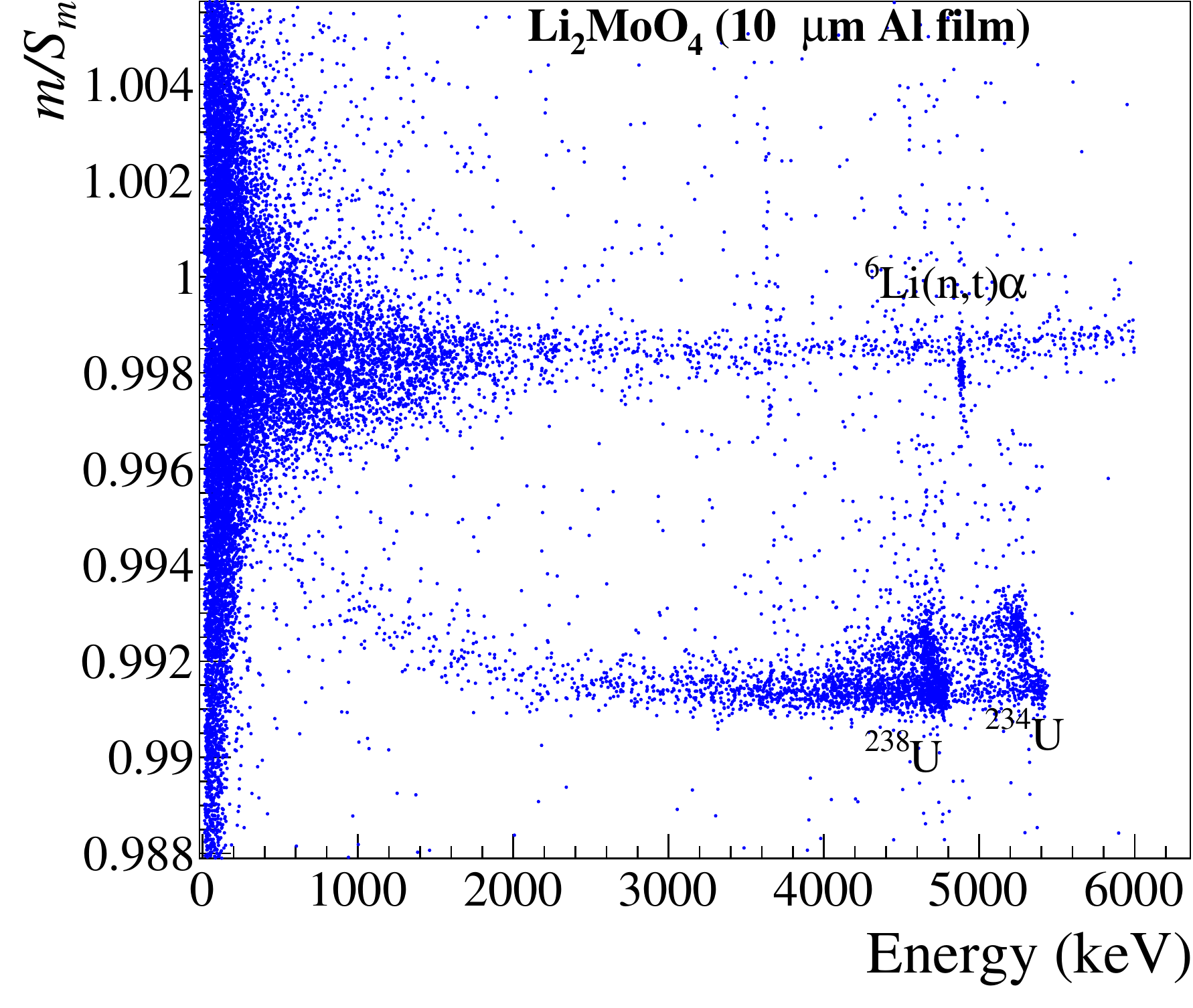}
\caption{\label{fig:surfsens} The parameter $m/S_m$ (see text) is plotted as a function of the energy (according to a $\gamma$ calibration) for a TeO$_2$ bolometer (Detector 2 in Table~\ref{tab:summary}) (left panel) and for a Li$_2$MoO$_4$ bolometer (Detector 1 in Table~\ref{tab:summary}) (right panel). The uranium-source $\alpha$ events are clearly discriminated. The main features of the plots are discussed in the text. In particular, the neutron capture events on $^{6}$Li are discussed in Section~\ref{sec:neutron}, while the shape of the populations related to the uranium $\alpha$ particles is discussed in Section~\ref{sec:rejalpha}. }
\end{figure}

Surface events induced by $\alpha$ particles can be safely rejected by PSD. We have tested several PSD parameters, which can provide surface/bulk event separation with different efficiencies. A good pulse-shape parameter is simply the pulse rise-time, which is the most intuitive one and easy to relate to a physical interpretation. Another efficient parameter is obtained in the following way. An average pulse, typically dominated by bulk events, is constructed. Let's define $A_i=A(t_i)$ the sampled amplitudes at the time $t_i$ of the average pulse $A(t)$ (with maximum amplitude normalized to 1), and $S_i=S(t_i)$ the corresponding sampled amplitudes of an individual signal $S(t)$ without normalization. The two pulses are synchronized at their maximum. If we plot $S_i$~vs.~$A_i$, we obtain a straight line crossing the origin and with slope equal to 1 for an individual signal identical to the average pulse. If we perform a linear fit of the data ($A_i$,$S_i$) for a generic signal, we obtain two parameters, the y-intercept $q$ and the slope $m$. The latter is an estimator of the individual-pulse amplitude and it almost coincides with that extracted with the optimal filter ($S_m$ ), if the signal has the same shape as that of the average pulse. (In practice, $m$ turns out to be slightly lower than $S_m$ in our detectors, at the per-mil level.) The y-intercept would be zero if, again, the average and the individual pulses had the same shape. It turns out that $m/S_m$ is a very sensitive pulse-shape parameter, and  we have extensively used it in the data analysis. We note that other parameters could be even more effective, but we have not performed a systematic study on the best way to account for pulse-shape differences yet.  

The discrimination can be appreciated in Fig.~\ref{fig:surfsens}, where the PSD parameter is $m/S_m$ for both crystals, and the Al-film thickness is 10~$\mu$m. The two lines of the uranium source, which irradiates the Al film, are clearly separated from the bulk events. The TeO$_2$ measurement (Fig.~\ref{fig:surfsens}, left panel) was performed in a run in which the polonium source was present and, as expected, it contributes to the bulk-event population. The tails exhibited by the polonium cluster are symmetric and are due to analysis issues related to the high activity of the source. In fact, the tails correspond to not rejected pile-up events on the trailing edge of the main $\alpha$ pulse, which affect the pulse shape without affecting significantly the pulse amplitude. The Li$_2$MoO$_4$ measurement (Fig.~\ref{fig:surfsens}, right panel) was performed in another run, with no polonium source. The neutron-capture events leak outside the bulk-event band towards low values of $m/S_m$ as they are partially reconstructed as surface events, as discussed in Section~\ref{sec:neutron}. The bulk events are mainly induced by the $\gamma$ activity below 2615 keV (limit of the natural $\gamma$ environmental spectrum) but they include also, at higher energy, the cosmic-muon interactions, which are not fully vetoed by the light detector. In fact, muons in general cross all the crystal and deliver a negligible fraction of their energy in the film or in its proximity. Below $\sim 2$~MeV, the width of the bulk event band tends to increase as the energy decreases. This is due both to a worsening of the signal-to-noise ratio and to pile-up effects which alter the signal shape. It is not clear whether the lowering of $m/S_m$ in the same energy region observed in surface $\alpha$ events is also due these spurious effects or to not-yet-identified physical reasons. The $\alpha$ and neutron-capture events are not at the correct energy positions (with the exception of the $^{210}$Po $\alpha$ line~\cite{Bellini:2010a}) because the detector response to heavy charged particles is different from that to $\gamma$'s in Li$_2$MoO$_4$~\cite{Bekker:2016a} and --- in addition --- the thermalization efficiency of close-to-film events is higher for both TeO$_2$ and Li$_2$MoO$_4$ and maximum for full absorption in the film (see Section~\ref{sec:NbSiNTD}). This mismatch does not affect the search for $0\nu2\beta$ decay, as the bulk region, which contains the potential signal, can be safely calibrated with medium-high-energy $\gamma$ sources. In addition, the efficiency of the surface-event discrimination is not decreased by the presence of two slightly different energy scales.

\begin{figure}[tbp]
\centering 
\includegraphics[width=0.7\textwidth]{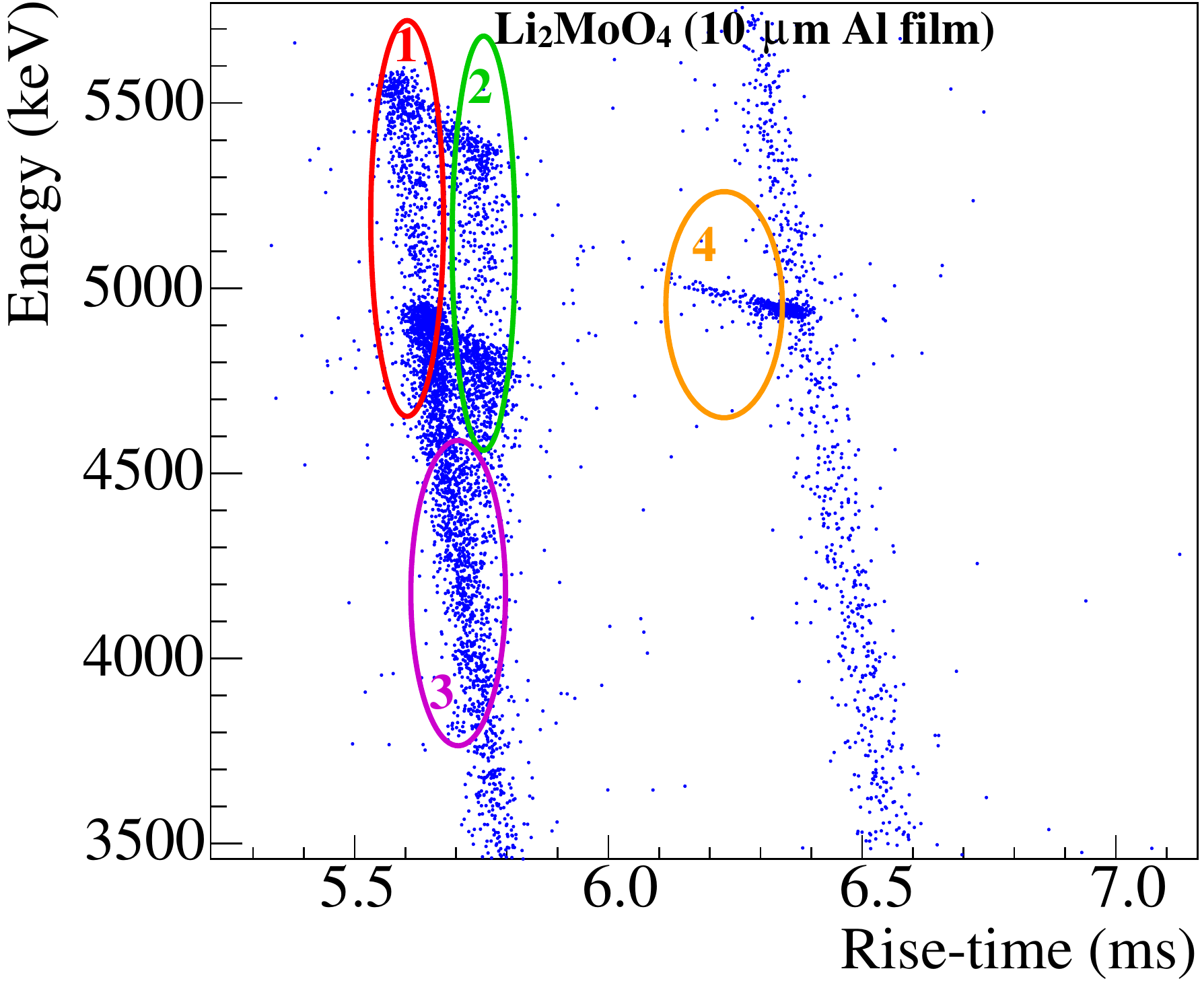}
\caption{\label{fig:risevsamp} Energy as a function of the rise-time for a Li$_2$MoO$_4$ bolometer (Detector 1 in Table~\ref{tab:summary}), with a 20$\times$20~mm side covered by the Al film and illuminated by an uranium source. The different populations evidenced in the four colored ovals are discussed in the text. As for the energy of the $\alpha$ and neutron-capture events, see caption of Fig.~\ref{fig:surfsens}.}
\end{figure}

\begin{figure}[tbp]
\centering 
\begin{overpic}[width=0.75\textwidth]{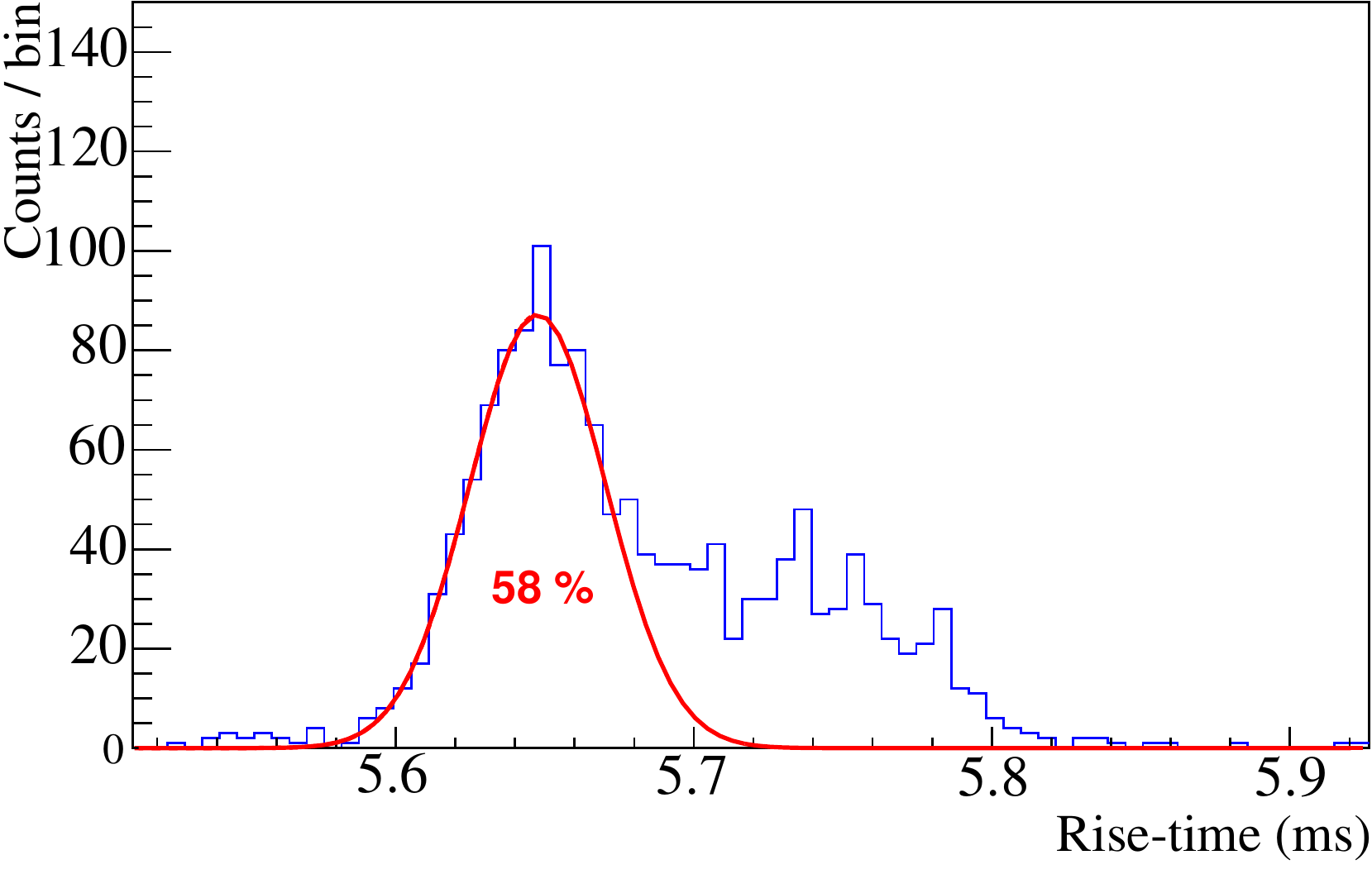}
   \put(50,28){\includegraphics[width=0.35\textwidth]{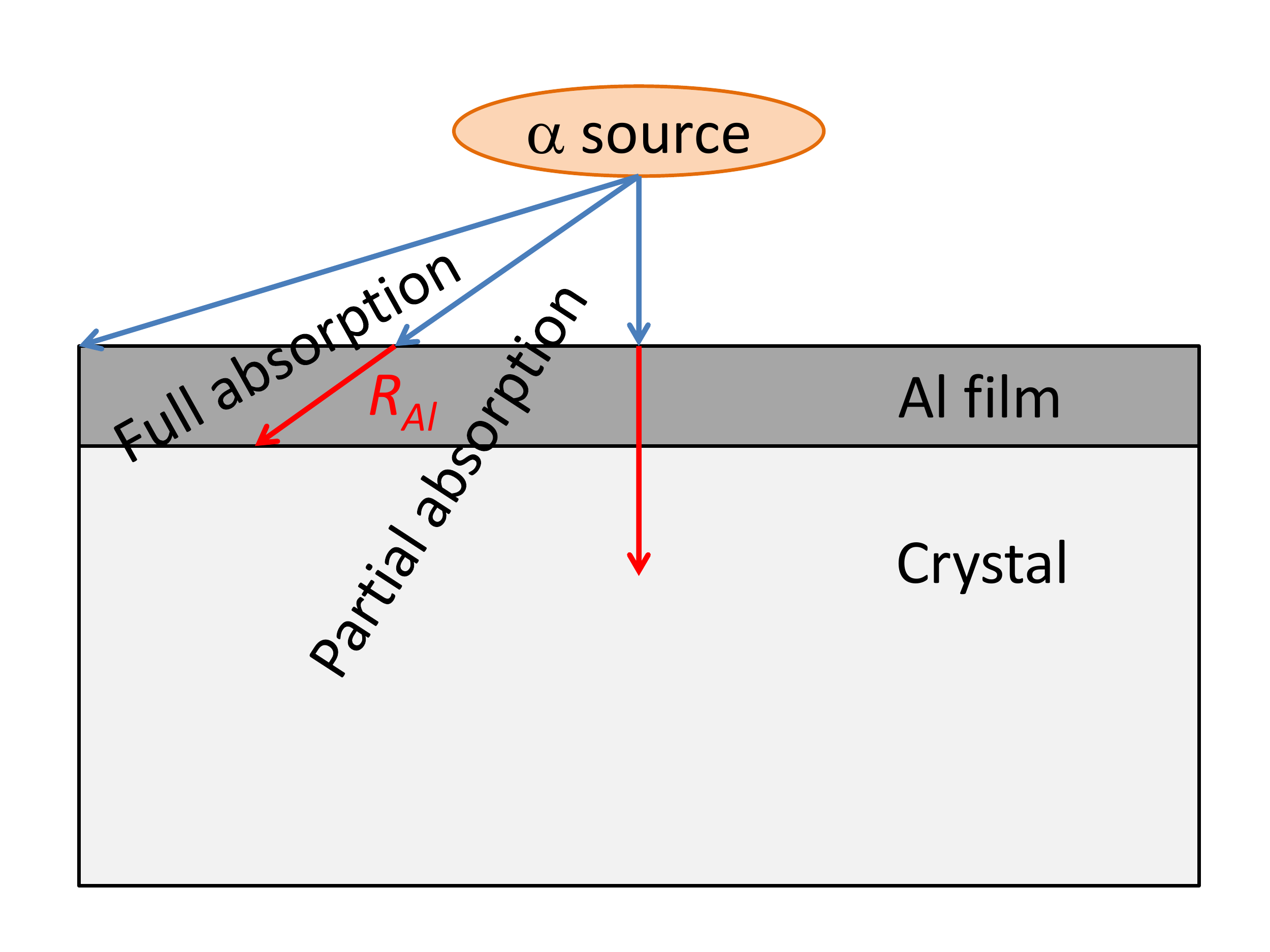}}
\end{overpic}
\caption{\label{fig:alfadistr} Rise-time distribution of signals corresponding to the $\sim 4.2$~MeV $\alpha$-particle from the uranium source. The fraction of $\alpha$ particles that deposit their full energy in the 10-$\mu$m-thick Al film coating the Li$_2$MoO$_4$ crystal is $\sim 58$\%, in agreement with a Monte Carlo simulation. In the inset, schematic geometry of the $\alpha$-particle absorption at the film-coated side of the crystal. $R_{Al}$ represents the range of $\alpha$ particles in aluminum.}
\end{figure}

We will comment now the event distribution in the $\alpha$-particle region. We use as a reference the measurement performed with the Li$_2$MoO$_4$ crystal coated with a 10-$\mu$m-thick Al film (Detector 1 in Table~\ref{tab:summary}). The pulse amplitude as a function of the rise-time is shown in Fig.~\ref{fig:risevsamp}. As anticipated above, surface events are characterized by a shorter rise-time. Part of the $\alpha$ particles are fully absorbed in aluminum (when they are emitted at a small angle with the surface), while the remaining part will share the deposited energy between the film and the crystal. In the inset of Fig.~\ref{fig:alfadistr}, the geometry of $\alpha$ absorption is depicted schematically. We stress that the film thickness is similar to the $\alpha$ range in aluminum: 16.8 $\mu$m and 20.0 $\mu$m at 4.2 MeV and 4.7 MeV respectively. This similarity affects strongly the distribution of the $\alpha$ events, as an important fraction of them corresponds to full absorption in the film. In Fig.~\ref{fig:risevsamp}, the region inside the red contour (n.~1) includes fully-film-absorbed $\alpha$ particles, emitted at a small angle with respect to surface. The region inside the green contour (n.~2) indicates mixed film-crystal energy depositions, for $\alpha$ particles entering the crystal with a direction almost normal to the surface. The ratio between the number of events in region n.~1 and that in region n.~2 is compatible with the geometry of the arrangement and the $\alpha$ particle ranges. It is expected that $\sim$~58\% of the $\alpha$ particles are fully absorbed in aluminum for the 4.2~MeV line according to a GEANT4-based~\cite{Agostinelli:2003a,Allison:2006a} Monte Carlo simulation, and this corresponds to what is estimated in the experimental distribution (Fig.~\ref{fig:alfadistr}). Finally, the region inside the violet contour (n.~3) is populated by energy-degraded  $\alpha$ particles that stop in the Al film, as expected since energy degradation requires that the particles exit at a small angle with the surface or come from the deepest part of the source.

It is interesting to note that the rise-time in Fig.~\ref{fig:risevsamp} is anti-correlated with the pulse amplitude for mono-chromatic energy depositions (the energy scale is set for bulk events using a $\gamma$ calibration). This can be easily explained by a qualitative interpretation of the phenomenon that we observe, which will be discussed in Section~\ref{sec:NbSiNTD}.

Fig.~\ref{fig:surfsens} shows a significant difference between the $\alpha$ distributions in TeO$_2$ and in \linebreak Li$_2$MoO$_4$. In the former case, the distribution has a larger spread. This suggests that the rejection capability in Li$_2$MoO$_4$ for full aluminum absorption is more similar to that for partial alu-minum absorption with respect to the TeO$_2$ case. (We do not have an interpretation of this behavior, which could be connected to the different high-energy phonon spectrum generated by $\alpha$ particles in the two materials.) As a consequence, we expect that a reduction of the film thickness should impact less the discrimination capability of Li$_2$MoO$_4$ with respect to that of TeO$_2$.

\begin{figure}[tbp]
\centering 
\includegraphics[width=0.75\textwidth]{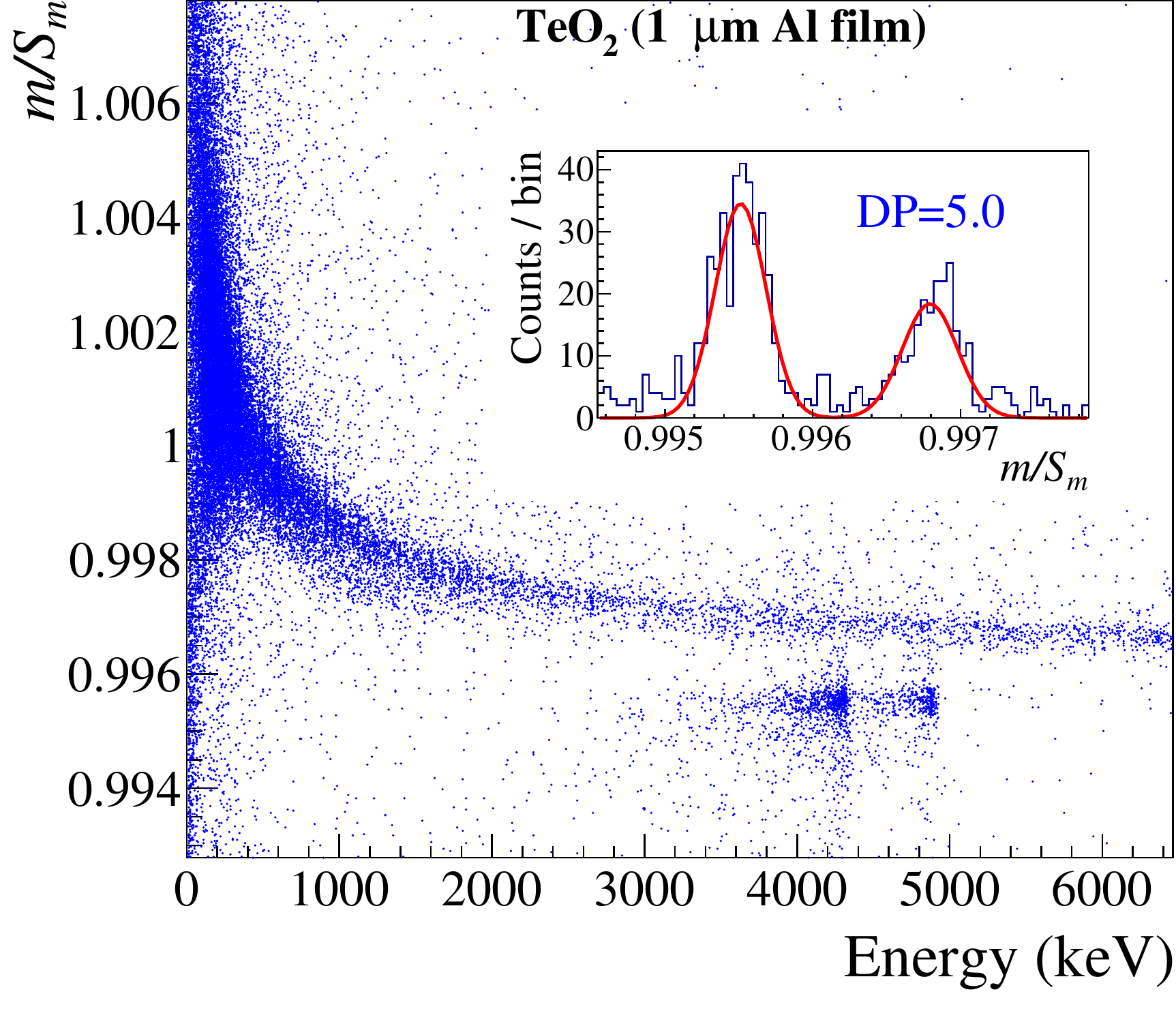}
\caption{\label{fig:onemicron} The parameter $m/S_m$ (see text) is plotted as a function of the energy for a TeO$_2$ bolometer (Detector 3 in Table~\ref{tab:summary}). The two lines of the uranium source, which irradiates the 1-$\mu$m-thick Al film, are clearly separated from bulk events. As for the energy of the $\alpha$ events, see caption of Fig.~\ref{fig:surfsens}. In the inset, Gaussian fits of the surface and bulk event distributions allow us to extract the discrimination power DP. The energy range of the selected events is 4400--5000 keV on a $\gamma$ calibration scale, corresponding to the higher-energy $\alpha$ line.}
\end{figure}

We have anyway shown that a film thickness of 1~$\mu$m is sufficient to perform discrimination at the required level in the case of TeO$_2$ (see Fig.~\ref{fig:onemicron}), using Detector 3 of Table~\ref{tab:summary}. The discrimination power decreases with respect to the 10~$\mu$m film thickness, but it remains excellent. In order to evaluate it, we have fitted by Gaussian functions the surface and bulk event distributions (see Fig.~\ref{fig:onemicron}, inset) to estimate their mean values ($\mu_{surface}$, $\mu_{bulk}$) and standard deviations ($\sigma_{surface}$, $\sigma_{bulk}$). After defining
\begin{equation}
{\rm DP} = \frac{\left| \mu_{surface} - \mu_{bulk} \right|}{\sqrt{\sigma^2_{surface}+\sigma^2_{bulk}}} \ ,
\end{equation}
as usually done for scintillating bolometers (with the pair $\alpha$-$\beta$ replaced by the pair $surface$-$bulk$), we have obtained an excellent ${\rm DP} = 5.0$, which is compatible with a rejection of surface events of 99.9\% with practically full acceptance of bulk events. This result is extremely important since for the 1-$\mu$m option the evaporation time is much shorter, the film adhesion is much better and the total mass of the material is decreased by one order of magnitude, lowering correspondingly the possible contribution of films to the radioactivity budget. For these reasons, we will consider 1-$\mu$m-thick Al films as the baseline solution for future CROSS detectors, and in particular for the medium-scale demonstrator described in Section~\ref{sec:prospect}. As argued above, DP (with 1-$\mu$m-thick Al coating) should be even better in case of Li$_2$MoO$_4$ crystals. 

Since the range of the uranium-source $\alpha$ particles in Al are in the 16--20 $\mu$m~range, it is clear that the fraction of fully-film-absorbed $\alpha$ particles is negligible for 1-$\mu$m-thick films. The achieved DP shows therefore that PSD works also for energies deposited up to few tens of $\mu$m directly in the TeO$_2$ crystals. It is therefore natural to wonder down to which distance from the film the surface events can be separated from the bulk ones. This is investigated in the next Section.

\subsection{Depth-dependence of surface sensitivity}
\label{sec:neutron}

\begin{figure}[tbp]
\centering 
\includegraphics[width=0.75\textwidth]{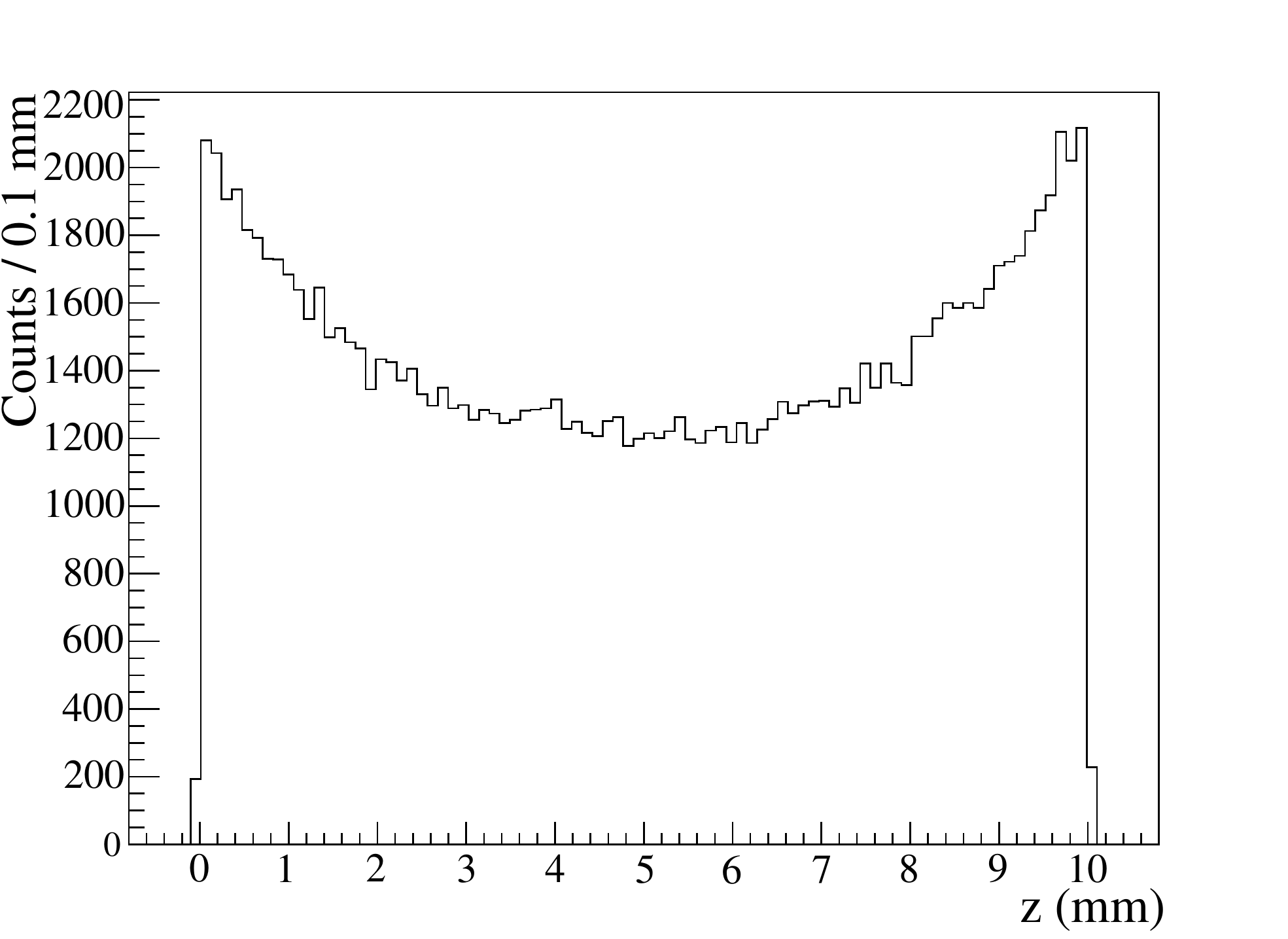}
\caption{\label{fig:impactpoint} Distribution of the distances of the neutron-capture sites from the Al film evaluated by a Monte Carlo simulation in a 20$\times$20$\times$10 mm Li$_2$MoO$_4$ detector (10$^6$ generated neutrons, 125468 neutron captures). The film is located at $z=0$. The neutron field is supposed to be isotropic and mono-energetic with neutron energy of 25~meV.}
\end{figure}

The Li$_2$MoO$_4$ detector with 10-$\mu$m-thick Al film (Detector~1 in Table~\ref{tab:summary}) provides an almost ideal tool to investigate sensitivity to shallow events down to a few millimeters below the coating. In fact, the isotope $^6$Li, which has a natural abundance of 7.8\%~\cite{Meija:2012a}, has a very high cross section for thermal neutron capture (940 barns)~\cite{Chadwick:2011a}. In this discussion we assume that the thermalized neutrons reaching the detector are at room temperature. In fact, there is no large low-temperature bath with low-mass elements in the vicinity of the detector, and the integrated thickness of the cold shields (mainly copper, not efficient in neutron moderation) is below 1~cm.  The thermal-neutron capture on $^6$Li is followed by the emission of a 2.05-MeV $\alpha$ particle and a 2.73-MeV triton, releasing a total kinetic energy of 4.78 MeV. The ranges in Li$_2$MoO$_4$ are respectively of 7~$\mu$m and 40~$\mu$m. Therefore, the energy released after a neutron capture is confined within $\sim 50$~$\mu$m around the interaction point. Since the absorption length of thermal neutrons in Li$_2$MoO$_4$ is $\sim~6.6$~mm, it is clear that neutron captures are distributed deeply in the crystal, but a significant fraction will occur close to the film. In addition, the energy deposition for each event can be --- in a first approximation --- considered as point-like with respect to the interaction-length scale. Therefore, ambient thermal neutrons will interact with high probability with the Li$_2$MoO$_4$ crystal and will sample the full volume of the detector with point-like energy depositions of the order of 4.8~MeV each. The distribution of the distance of neutron-capture site from the film has been evaluated with a semi-analytic calculation and with a Monte Carlo~\cite{Agostinelli:2003a,Allison:2006a} (in the hypothesis of isotropic neutron flow with neutron energy of 25~meV), providing very similar results (see Fig.~\ref{fig:impactpoint}). We stress that an important contribution to the distribution of the impact points come from neutrons that reach the detector from the sides that are not covered.

\begin{figure}[tbp]
\centering 
\begin{overpic}[width=0.75\textwidth]{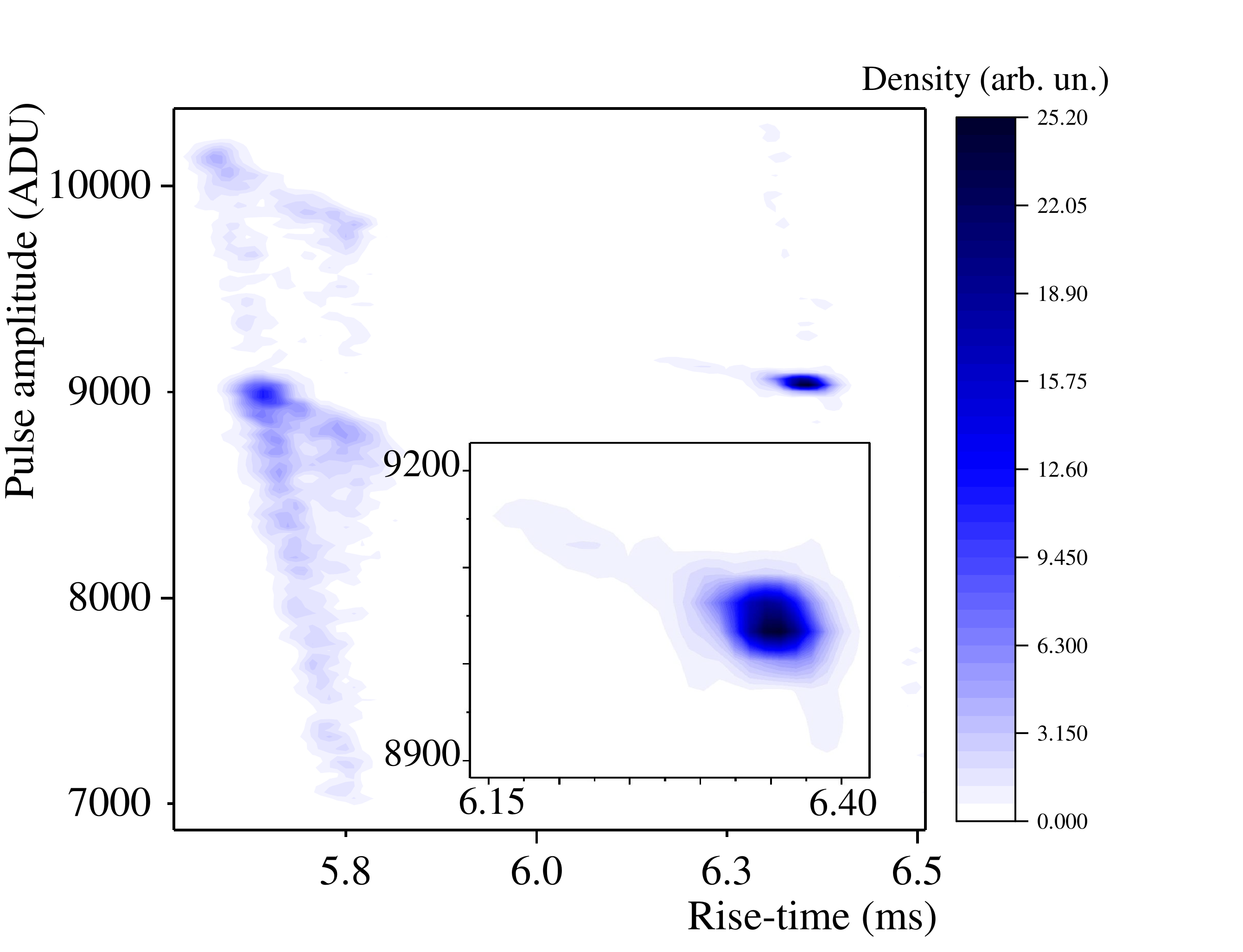}
\put(32,50){\small{$\longleftarrow$~$\alpha$ events}}
\put(33,44){\small{neutron events $\longrightarrow$}}
\end{overpic}
\caption{\label{fig:density} In this density plot, pulse amplitude as a function of the rise-time is shown for pulses acquired with Detector~1 of Table~\ref{tab:summary}. The $\alpha$-like events are selected with the light detector. Therefore, only events induced by the uranium $\alpha$ source and those corresponding to neutron captures on $^6$Li are contained in the plot. A zoom of the neutron-capture population is shown in the inset.}
\end{figure}

After these considerations, we refer now again to Fig.~\ref{fig:risevsamp}. It is clear that the neutron-capture events, in spite of the very long neutron absorption length with respect to the film thickness, do not lie entirely in the bulk event band, occupied mainly by cosmic-ray events in this energy region. They present a significant tail towards the surface-event region, as visible in Fig.~\ref{fig:risevsamp} (region n.~4, inside the orange contour). The distribution of the rise-times can be studied in detail by isolating the neutron-capture and $\alpha$ events using the light detector, thanks to their lower light yield for the same thermal energy~\cite{Pirro:2006a}. The selection of $\alpha$ and neutron events is shown in the density plot of Fig.~\ref{fig:density}. 

\begin{figure}[tbp]
\centering 
\includegraphics[width=0.75\textwidth]{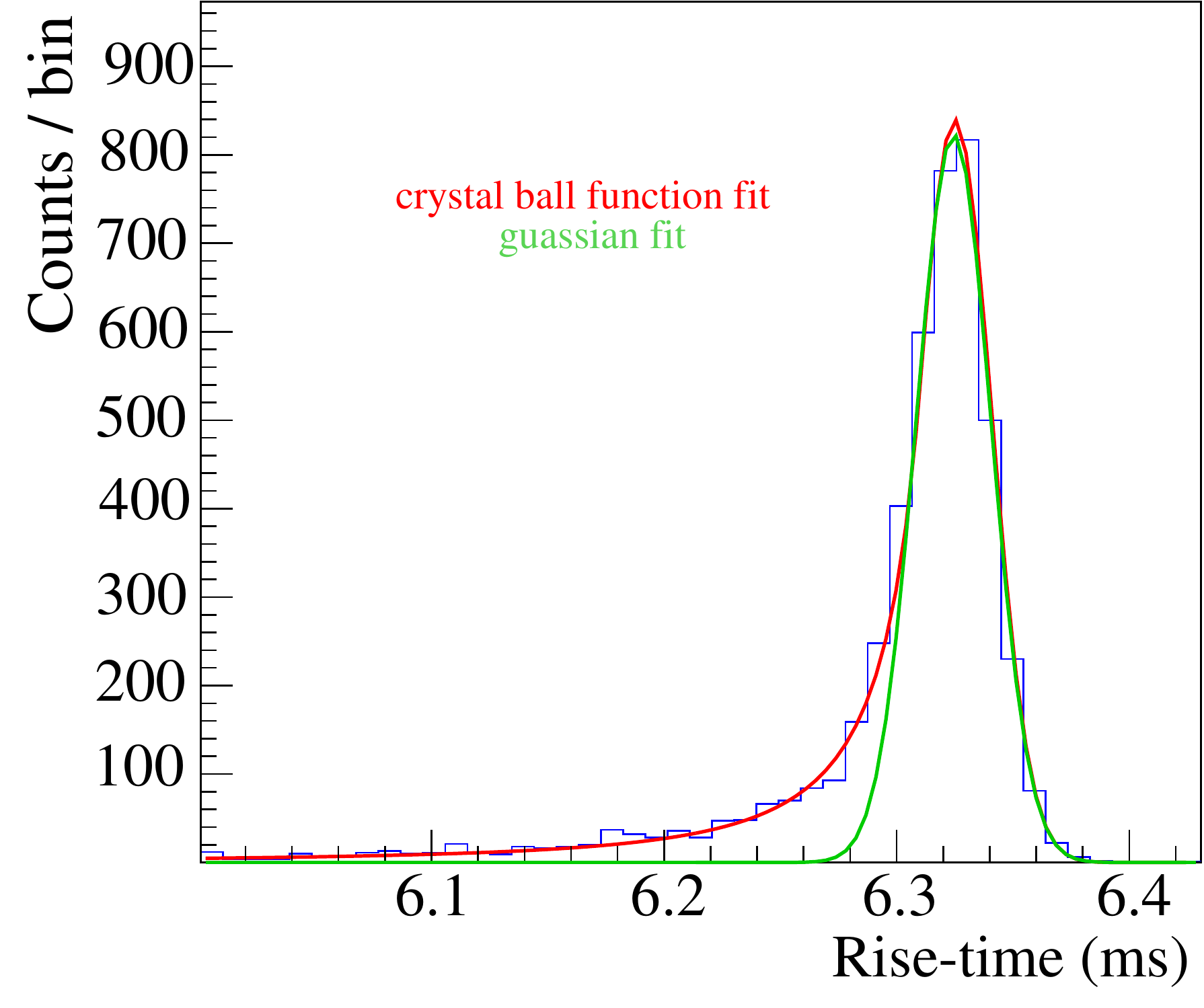}
\caption{\label{fig:risetime} Rise-time distribution for neutron-capture events collected by Detector 1 of Table~\ref{tab:summary}. The distribution is fitted with a Crystal Ball function. The tail at faster rise-time values contains close-to-Al-film events, while the main Gaussian peak contains bulk events.}
\end{figure}

It is now possible to correlate the rise-time distribution with the impact-point distribution discussed above. Fig.~\ref{fig:risetime} shows the rise-time distribution of all the neutron-capture events only (inset of Fig.~\ref{fig:density}), covering a rise-time interval of about $5.9-6.4$~ms. The Gaussian peak centered at long rise-time values ($\sim 6.32$~ms) contains the bulk events. The tail at short rise-times contains the close-to-film events, as we know from the $\alpha$-particle behavior. We have fitted the distribution with a Crystal Ball function (Fig.~\ref{fig:risetime}) and subtracted the Gaussian part from the full distribution, isolating in this way the tail component $S(t_R)$ containing close-to-film events. We have then constructed the ratio:
\begin{equation}
\label{eq:rise}
R_R(t_R) = \frac {\int_{t_{R_{min}}}^{t_R} dt'_R S(t'_R)} {S_R} ,
\end{equation}
which provides the fraction of surface events up to a rise-time $t_R$ with respect to the total number of neutron events (both surface and bulk) $S_R$, obtained by integrating over the entire distribution. Of course, Eq.~({\ref{eq:rise}}) is significant for $t_R < 6.3$~ms. We consider then the distribution $C(z)$ of the neutron-capture events in terms of $z$, depicted in Fig.~\ref{fig:impactpoint}, and construct the ratio:
\begin{equation}
\label{eq:z}
R_C(z) = \frac {\int_0^z dz' C(z')} {S_z} ,
\end{equation}
which provides the fraction of neutron-capture events occurring up to a distance $z$ from the Al film with respect to the total number of neutron events $S_z$.

\begin{figure}[tbp]
\centering 
\includegraphics[width=0.75\textwidth]{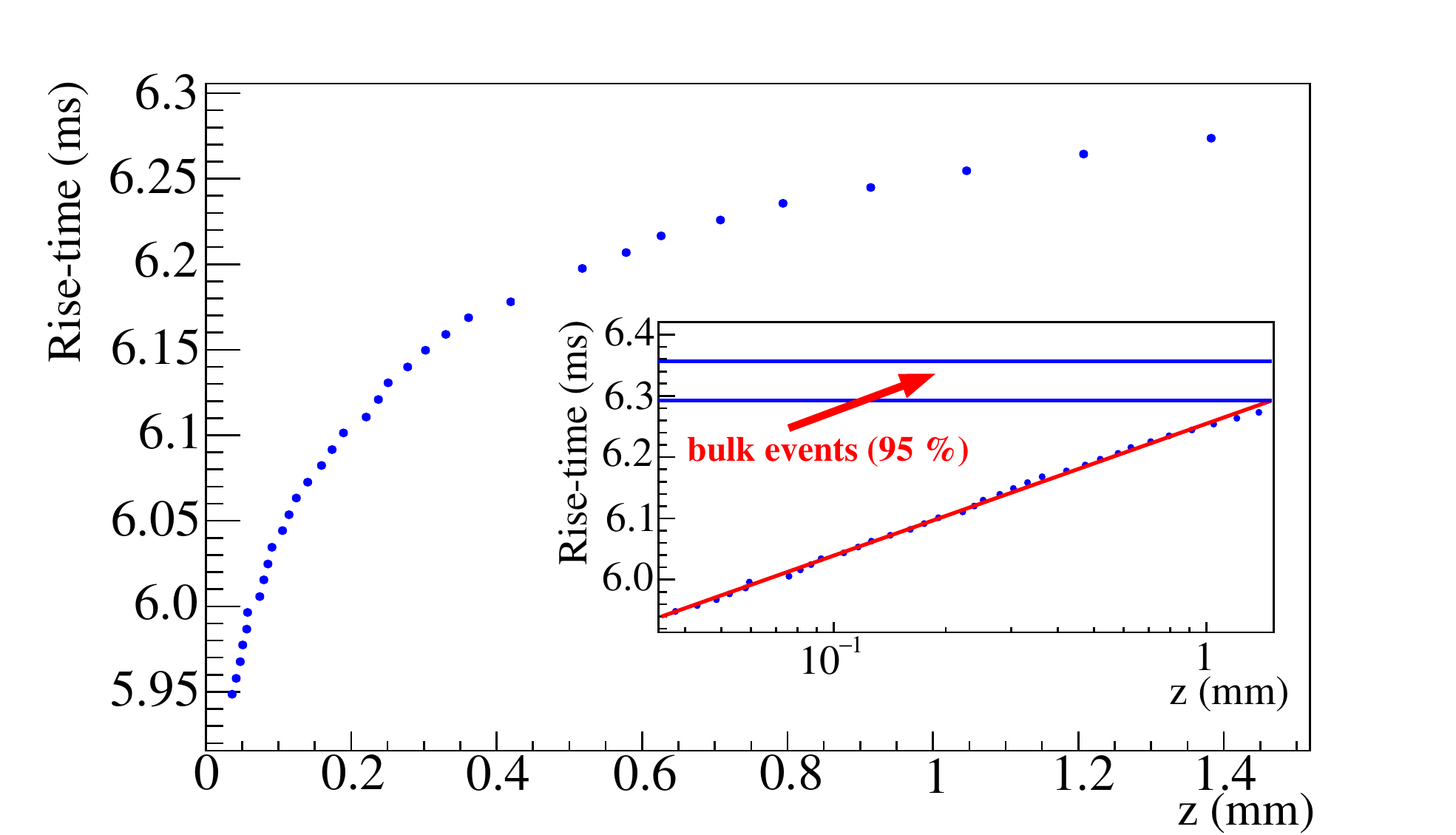}
\caption{\label{fig:rtvsdp} Pulse rise-time as a function of the depth of the neutron impact point for data collected with Detector 1 of Table~\ref{tab:summary}. In the inset, the same relationship in semi-log scale. The band containing 95\% of the bulk events is also shown. With the current rise-time resolution, surface events can be identified down to a distance of about 1.4~mm from the Al film.}
\end{figure}
  
We make now the assumption that the rise-time depends only on the distance $z$ of the neutron-capture site from the Al film, neglecting therefore edge effects. Using Eqs.~(\ref{eq:rise}) and (\ref{eq:z}), the relationship between $t_R$ and $z$ can then be determined by stating that a signal with a rise-time $\overline{t}_R$ (with $\overline{t}_R < 6.3$~ms) identifies a neutron-capture event occurred at a distance $\overline{z}$ from the Al film, such that
\begin{equation}
\label{eq:eqratio}
R_R(\overline{t}_R) = R_C(\overline{z}) .
\end{equation}
Fig.~\ref{fig:rtvsdp} shows the relationship provided by Eq.~(\ref{eq:eqratio}). Close-to-film events can be discriminated down to a depth of about 1.4~mm. Above this value, rise-time resolution (of the order of $\sigma = 20$~$\mu$s) does not allow us to separate surface from bulk events. In the inset of Fig.~\ref{fig:rtvsdp}, one can appreciate that the relationship between rise-time and depth is approximately linear when depth is plotted in logarithmic scale. The straight line corresponds to a phenomenological fit and is drawn to guide the eye. We have no phonon-physics model capable of explaining quantitatively this behavior yet. These data however are very useful to test future first-principle description of the observed mechanism. 

The depth-extended PSD capability of our technique --- down to 1--2~mm from the Al film --- is promising for $\beta$-surface-event discrimination, given the millimeter range of $\sim$~MeV~energy $\beta$ electrons. However, the $\beta$ component of the uranium source has not been detected so far. In fact, no low-energy component in the range $0-2.2$~MeV is appreciable in Fig.~\ref{fig:surfsens} in the surface-event region, with the exception of the aforementioned energy-degraded $\alpha$ population. This may be due to the fact that $\beta$ electrons lose energy with continuity along their millimetric path and the deposited-energy density is not high enough at the end of their range, in contrast with the neutron-capture events that are point-like with respect to a millimeter length scale. 

\subsection{Comparison of responses from athermal and thermal phonon sensors}
\label{sec:NbSiNTD}

In all the tests made with NTD Ge thermistors, surface events have a shorter rise-time with respect to bulk events. As argued above, this is in contrast with the expectations (Section~\ref{sec:surface-sensitivity}), which however were assuming an enhanced athermal-phonon sensitivity of the employed sensor. In order to clarify this issue, we have operated a TeO$_2$ crystal (20$\times$20$\times$5~mm size) equipped simultaneously with an NTD Ge thermistor and two NbSi films (Detector 4 in Table~\ref{tab:summary}). The NTD Ge thermistor is sensitive mainly to the thermal component of the signal due to its intrinsic slowness and the glue interface, while the two large-area (14$\times$14~mm) NbSi films are deposited directly on the crystal, each on a 20$\times$20~mm side. This makes these sensors highly sensitive to the prompt athermal component of the phonon population produced by the impinging particle. (In the following, we will refer to just one NbSi film that we have selected for its better sensitivity.) The Nb fraction in the 0.65-nm-thick film is 8.65\%. With these parameters, the film resistivity as a function of the temperature has a behavior similar to that exhibited by NTD Ge thermistors, featuring an exponential increase of the resistance as the temperature decreases~\cite{Dumoulin:1993a}. The film can be read out by the same electronics used for the NTD Ge thermistors, but the optimal resistances (several M$\Omega$'s) are achieved in the 30--40~mK temperature range, because of the extremely high impedance of the sensor. The NTD Ge thermistor, glued on a 20$\times$20~mm surface in the part not covered by the film, was selected in order to have resistances similar to that of the film in the same temperature range. A description of the NbSi-film-production procedure is described in Ref.~\cite{Nones:2012a}.

\begin{figure}[tbp]
\centering 
\includegraphics[width=0.75\textwidth]{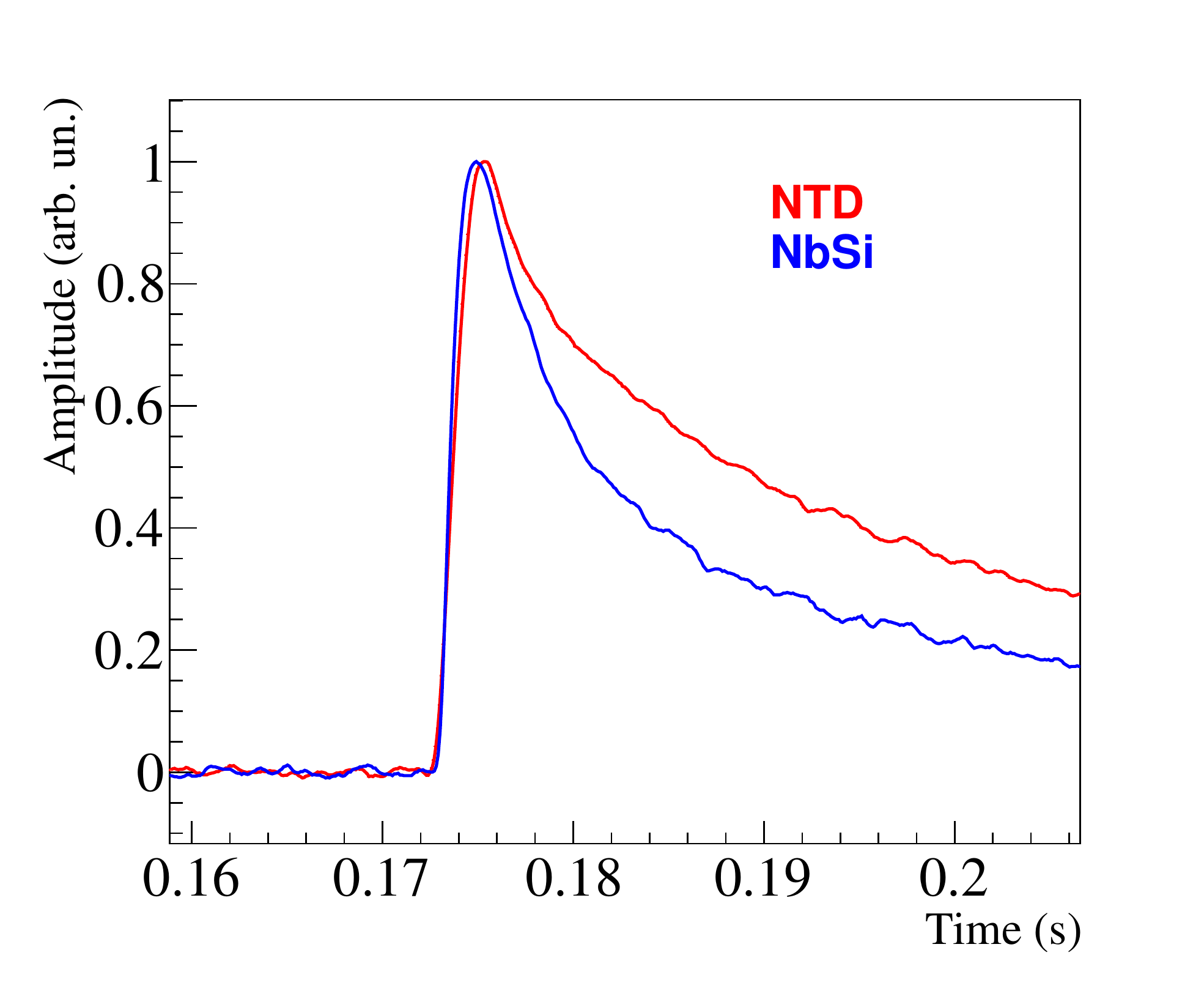}
\caption{\label{fig:NTDNbSicomparison} Comparison between normalized pulses provided by the NbSi sensor and by the NTD Ge thermistor for the same bulk event, collected with Detector 4 of Table~\ref{tab:summary}.}
\end{figure}

A 10-$\mu$m-thick Al film was evaporated on a 20$\times$5~mm side of the crystal. An uranium source irradiated the coated surface in order to generate the surface events to be discriminated, and a source of the same type was directly deposited on the opposite side of the crystal, providing bulk-like events (it was unpractical to use the polonium source in this particular detector configuration). The detector was operated at $\sim 30$~mK, with the two aforementioned phonon sensors working simultaneously.\footnote{To our knowledge, this is the first time that a composite bolometer has functioned with a double readout providing separated sensitivity to athermal and thermal phonons for the same event.} Fig.~\ref{fig:NTDNbSicomparison} shows a comparison of pulse shapes related to the same bulk event detected by the two sensors. The faster time development of the NbSi pulse is due to its important athermal-phonon component.

\begin{figure}[tbp]
\centering 
\includegraphics[width=0.49\textwidth]{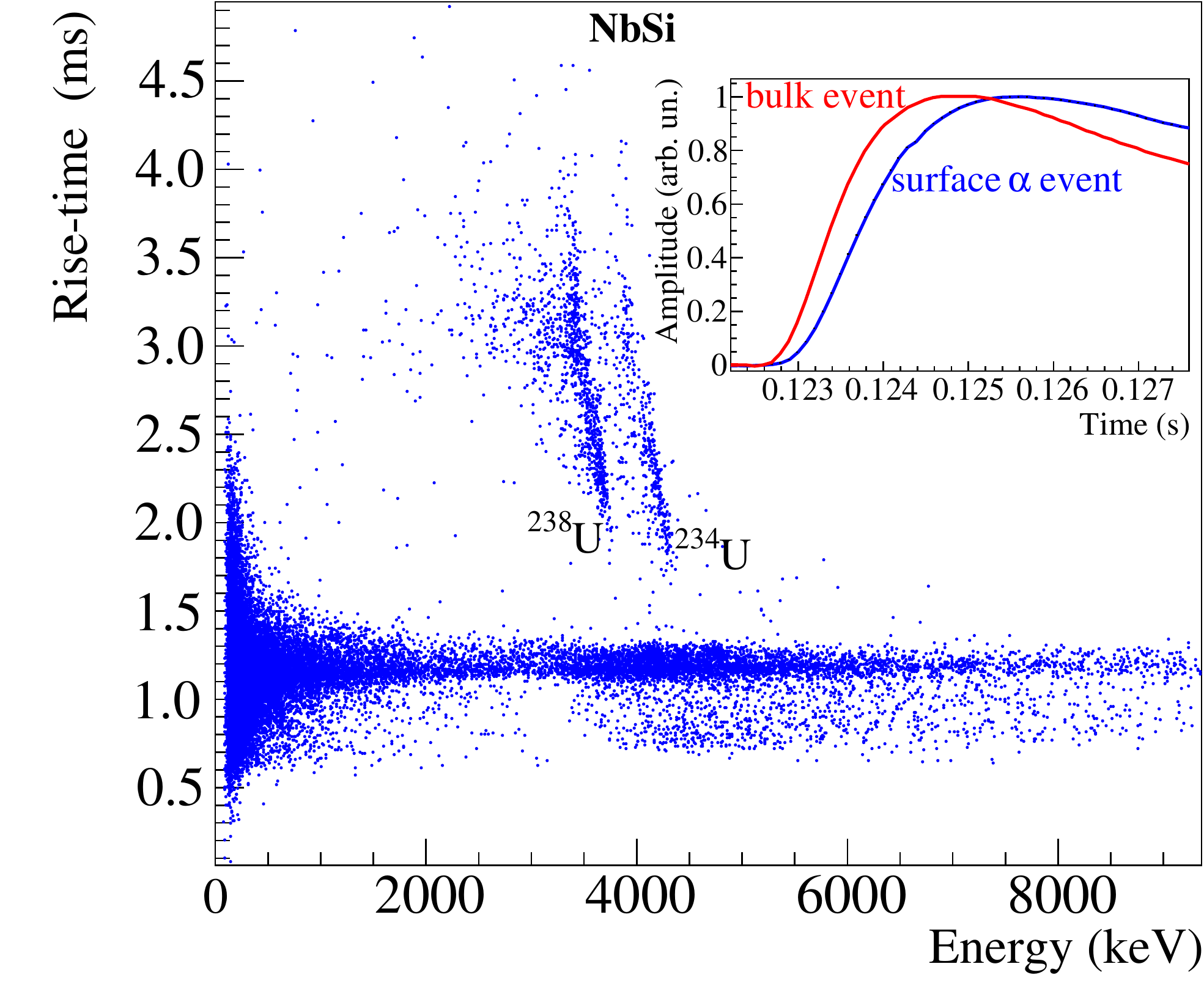}
\includegraphics[width=0.49\textwidth]{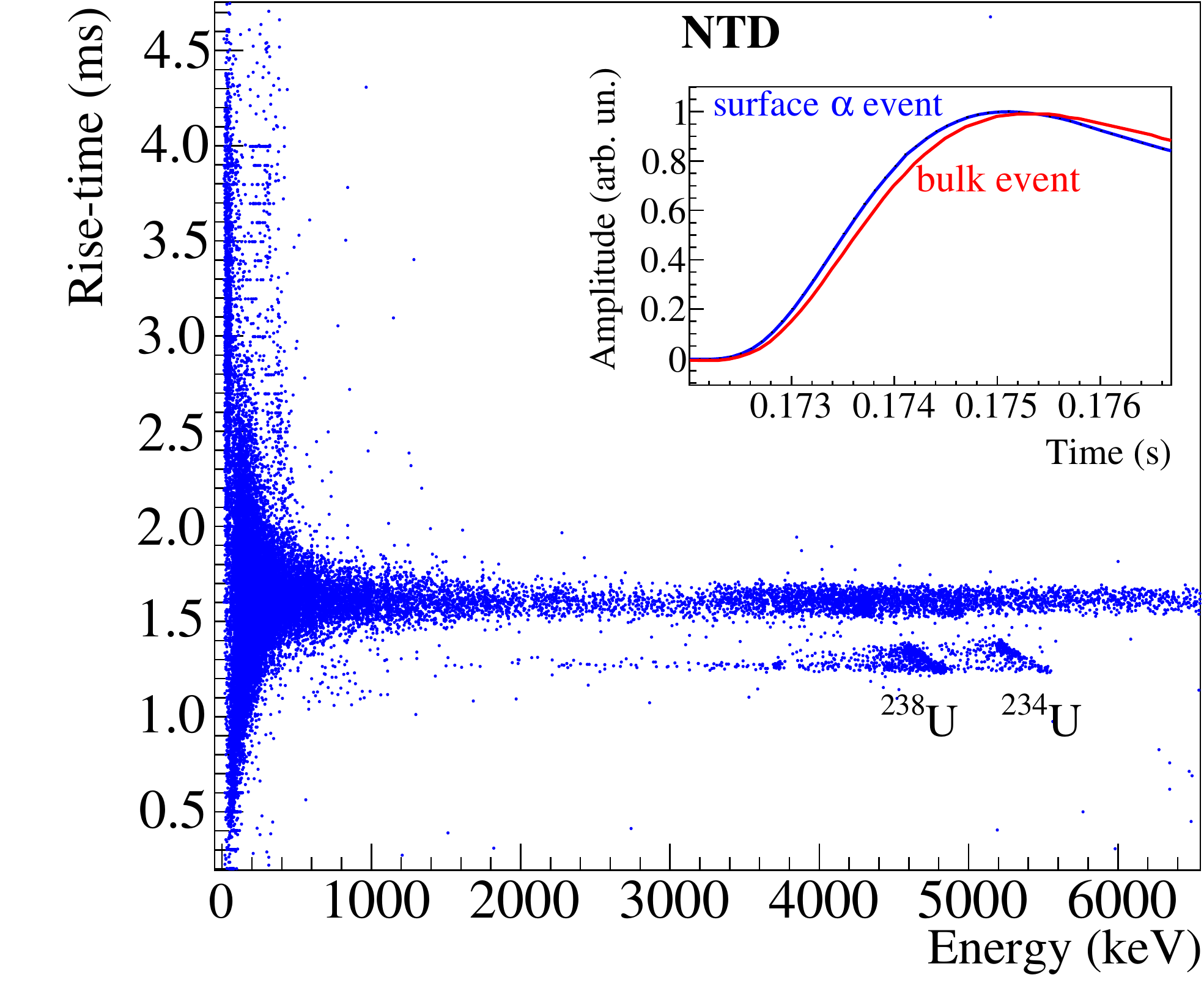}
\caption{\label{fig:NTDNbSiplots} The same set of events acquired by Detector 4 of Table~\ref{tab:summary} represented with rise-time vs. pulse-amplitude plots for NbSi-film signals (left panel) and NTD-Ge-thermistor signals (right panel). The inverted behaviour of the two sensors in terms of rise-time of surface events is apparent. As for the energy of the $\alpha$ events, see caption of Fig.~\ref{fig:surfsens}. In the insets, a direct comparison of signal rise-times for bulk and surface events is shown, both for NbSi sensor (left panel) and NTD Ge thermistor (right panel).}
\end{figure}

In Fig.~\ref{fig:NTDNbSiplots}, we report the amplitude as a function of the rise-time for the NbSi sensor (left panel) and for the NTD Ge thermistor (right panel). The insets show details of the rise-times of surface and bulk events for the two sensor types. The rise-time behavior is opposite in the two cases. It is confirmed that surface events are faster with the NTD Ge technique, as already observed in the results discussed in Section~\ref{sec:rejalpha}. On the contrary, surface events are slower when detected by the NbSi sensor, according to the arguments exposed in Section~\ref{sec:surface-sensitivity} and the results of Ref.~\cite{Nones:2012a}. In fact, these arguments hold when referred to the athermal phonon component of the signal. In case of NTD Ge thermistors, we have to consider that these slow devices are sensitive mainly to thermal phonons. The time required to initial high-energy phonons to degrade to thermal phonons is long (on the millisecond scale at least). If however a significant fraction of them is absorbed by the film, as happens in a surface event, the new phonon population that is re-injected from the film to the crystal after quasi-particle recombination has a much lower energy ($\sim 1.2$~K, superconductive gap of Al). This energy down conversion is accelerated with respect to what happens in bulk events, providing a faster signal for surface events (see Fig.~\ref{fig:phonon-scheme}, right panel). In short, the Al film stores and delays the release of energy associated to athermal phonons, but accelerates thermalization as it reduces faster the average phonon energy. This interpretation is consistent with the observed anti-correlation between rise-time and pulse amplitude for a fixed energy deposition (Fig.~\ref{fig:risevsamp}). In fact, the faster and more effective thermalization provided by the film enhances the amplitude of the thermal signal. 

In future tests, the NbSi option can provide a further mean to enhance the surface sensitivity in the CROSS technology. The simultaneous use of the two sensor types and their different time responses will in fact provide a redundancy in the identification of surface events that can be an asset for background rejection. Furthermore, the NbSi-thin-film sensors can improve the rejection of the background induced by random coincidences of the $2\nu2\beta$ events, as pointed out in Section~\ref{sec:isotope}.

\section{CROSS prospects: medium scale demonstrator and large-scale applications}
\label{sec:prospect}

A dilution refrigerator with an experimental volume of about $\sim 150$~l was recently (April 2019) installed in the Canfranc underground laboratory (Spain) in the framework of the CROSS project. This machine can provide a base temperature of $\sim 10$~mK and is optimized for the operation of macro-bolometers. A muon veto will be set up by covering with scintillating elements the walls of the hut housing the cryostat. We plan to install a first demonstrator of the CROSS technology in this cryogenic set-up. It will consist of 32 Li$_2$MoO$_4$ crystals grown with molybdenum enriched in $^{100}$Mo at $> 95$\% level. Each crystal is a cube with 45~mm side, for a mass of 0.28~kg. The crystals are already available and have been produced according to a protocol developed in the LUMINEU experiment, which ensures excellent bolometric performance and high radio-purity~\cite{Armengaud:2017a,Poda:2017b,Armengaud:2019a,Berge:2014a}. The LUMINEU technology ensures $\sim $~5~keV FWHM energy resolution and a background level inferior to $10^{-4}$ counts/(keV~kg~y) in the region of interest around 3034~keV as far as the crystal contribution is concerned~\cite{Artusa:2014a,Beeman:2012a}. The crystals will be arranged in a tower of eight floors with four crystals each, with almost full visibility between adjacent elements, in order to improve the background rejection through anti-coincidences. The total $^{100}$Mo mass will be of 4.7~kg, corresponding to $2.9 \times 10^{25}$ $^{100}$Mo nuclei.

The techniques described above for the rejection of surface events will be extended to these large-volume crystals. The evaporator installed at CSNSM will be upgraded with mechanical supports in order that ultra-pure Al films can be deposited on all the large surfaces of the cubic crystals, which will be fully coated. The thickness of the films will be 1~$\mu$m. The protocol described in Section~\ref{sec:set-up} will be adopted. We will assume a discrimination power ${\rm DP} = 5.0$, as measured in the test described in Section~\ref{sec:rejalpha} for this film thickness and shown in the inset of Fig.~\ref{fig:onemicron}. This DP value corresponds to a rejection factor of $\alpha$ surface radioactivity higher than 99.9\% with an acceptance higher than 99\% for bulk events. This figure was obtained in the interval 4400--5000~keV, but it can be safely extrapolated down to 3000~keV since no enlargement of the bulk-event band is appreciable down to this energy, as visible in Fig.~\ref{fig:onemicron}. We expect that $\beta$ surface events are not cut by the adopted PSD method, as unfortunately our tests did not show the capability of selecting them. Therefore, we will assume full $\beta$ sensitivity even after the cut for surface events. Of course, we have to prove that the surface events can be safely identified also in the large samples envisioned in the CROSS demonstrator. This will be investigated in preliminary dedicated tests. The mechanism of surface-event identification should not however be affected by the crystal size. No light detector is foreseen. 

In order to evaluate the sensitivity of this demonstrator, we have to make assumptions on the background level. We do not have developed a full background model of our set-up yet. However, considering that our cryostat has been fabricated with a careful selection of the materials and that the shielding was designed in order to achieve very low background in the experimental volume, we will assume a background level in the range $10^{-2} - 10^{-3}$ counts/(keV~kg~y). In the ROI for $0\nu2\beta$ decay of $^{100}$Mo, these values have been already demonstrated in a similar-size bolometric set-up placed underground (CUORE-0~\cite{Alfonso:2015a} and CUPID-0~\cite{Azzolini:2018a} experiments). In particular, the capability of rejecting $\alpha$ surface radioactivity has shown that it is possible to approach the $10^{-3}$ counts/(keV~kg~y) value~\cite{Azzolini:2019a}. 

The results on the sensitivity to $0\nu2\beta$ decay are reported in Table~\ref{tab:demonstrator}. The limits on the expected number of counts in the ROI have been calculated according to the Feldman-and-Cousins confident-belt construction~\cite{Feldman:1998a}. We notice that the 2-year sensitivity to the effective Majorana neutrino mass $m_{\beta\beta}$, considering computations of nuclear matrix elements taken from different models~\cite{Simkovic:2013a,Hyvarinen:2015a,Barea:2015a,Vaquero:2013a,Yao:2014a}, is in the range $\sim$~100--200~meV. CROSS can therefore compete with all the present searches but the KamLAND-Zen experiment~\cite{Gando:2016a}, now leading the field. The 5-year sensitivity, in case of favorable background conditions ($10^{-3}$ counts/(keV~kg~y)), is practically equivalent to the current limit of KamLAND-Zen. This shows the potential of the CROSS technology, considering the low size and the mere demonstration character of our proposed set-up.

\begin{table}[tbp]
\centering
\begin{tabular}{|cccccc|}
\hline
Background & Live & Background & Feldman-Cousins & Half-life & $m_{\beta \beta}$ \\
level & time & counts & count limit & limit $[$y$]$& limits $[$meV$]$ \\
$[$counts/(keV kg y)$]$ & $[$y$]$ & in ROI  & (90\% c.l.) & (90\% c.l.) & (90\% c.l.)  \\
\hline 
$10^{-2}$ & 2 & 1.4 & 3.6 & $8.5\times10^{24}$ & 124--222 \\
$10^{-3}$ & 2 & 0.14 & 2.5 & $1.2\times10^{25}$ & 103--185 \\
$10^{-2}$ & 5 & 3.6 & 4.6 & $1.7\times10^{25}$ & 88--159 \\
$10^{-3}$ & 5 & 0.36 & 2.7 & $2.8\times10^{25}$ & 68--122 \\
\hline
\end{tabular}
\caption{\label{tab:demonstrator} Sensitivity of the CROSS demonstrator to be operated in the cryogenic facility of the Canfranc underground laboratory, under two different hypotheses of background level. The region of interest is 8 keV. The efficiency is 75\%, compatible with the crystal size of each element. The interval of limits on the effective Majorana neutrino mass $m_{\beta \beta}$ accounts for a recent compilation of nuclear-matrix-element calculations.}
\end{table}

We foresee also a $^{130}$Te~section in the CROSS demonstrator, but the design of this part is less advanced and the study of a protocol for the development of radio-pure crystals of TeO$_2$ grown with enriched tellurium is still under refinement, on the basis of the results described in Ref.~\cite{Artusa:2017a}. 

As discussed in Section~\ref{sec:background}, the results achieved by CUORE and the related preliminary background model shows that a cryogenic infrastructure suitable for housing $\sim 1000$ large-mass bolometers is compatible with a background level of only $10^{-4}$ counts/(keV~kg~y) at $\sim 3$~MeV, i.e. in the vicinity of the $^{100}$Mo $0\nu2\beta$ signal. That low background can be achieved if and only if the $\alpha$ surface radioactivity is rejected at 99.9\% level. The baseline solution for this background reduction is the scintillating-bolometer technology, as proposed in the CUPID project~\cite{Wang:2015a,CUPID:2019a}. CROSS can provide an adequate background abatement with a major simplification of the detector structure and of the readout. When applied to large arrays containing $0\nu2\beta$-candidate masses of the order of hundreds of kilograms, the CROSS technology has the potential to explore fully the inverted-ordering region and to detect $0\nu2\beta$ decay even in case of direct ordering, provided that the lightest neutrino mass is higher than $\sim$~10--20~meV. 

\section{Conclusions}
\label{sec:conclusions}

In this paper, we have presented a new bolometric technology capable of setting the grounds for a future experiment searching for $0\nu2\beta$ decay with a background level of only 0.5 counts/(ton y) and compatible with ton-scale setups. These features will enable the search for lepton number violation with unprecedented sensitivity, penetrating in prospect the direct-ordering band of the neutrino masses in a region of the parameter space where it does not overlap with the inverted-ordering band. The key idea in CROSS is to reject surface events --- a dominant background source in bolometric experiments --- by pulse-shape discrimination, obtained by exploiting solid-state-physics phenomena in superconductors. We have shown indeed in a series of prototypes that ultra-pure superconductive Al films deposited on the bolometer surfaces act successfully as pulse-shape modifiers, both with fast and slow phonon sensors, although the mechanisms are different in the two cases. This method will allow us to get rid of the light detectors used up to now to discriminate surface $\alpha$ particles, simplifying a lot the bolometric structure. The intrinsic modularity and the simplicity of the read-out will make the CROSS technology easily expandable. The current program is focused on an intermediate experiment to be installed underground in the Canfranc laboratory in an existing dedicated facility. Its purpose is to test the technique with high statistics and to prove the stability and the reproducibility of the CROSS methods, but at the same time this experiment will be competitive in the $0\nu2\beta$-decay international context.

\acknowledgments
The project CROSS is funded by the European Research Council (ERC) under the \linebreak European-Union Horizon 2020 program (H2020/2014-2020) with the ERC Advanced Grant no. 742345 (ERC-2016-ADG). The ITEP group was supported by Russian Scientific Foundation (grant No. 18-12-00003). The PhD fellowship of H. Khalife has been partially funded by the P2IO LabEx (ANR-10-LABX-0038) managed by the Agence Nationale de la Recherche (France) in the framework of the 2017 P2IO doctoral call. M.M.~Zarytskyy was supported in part by the project ``Investigations of rare nuclear processes'' of the program of the National Academy of Sciences of Ukraine ``Laboratory of young scientists''.

\bibliographystyle{JHEP}
\bibliography{CROSS-article-1.bib}








\end{document}